\documentclass[iop,twocolumn,tighten,numberedappendix,twocolappendix,revtex4]{emulateapj}
\usepackage{graphicx}
\usepackage{amsmath}
\usepackage{amssymb}
\usepackage{lipsum}
\usepackage{xcolor}

\shorttitle{34GHz Deep Field: Continuum properties}
\shortauthors{Algera et al.}

\begin{document}

\title{COLDz: Deep 34\,GHz Continuum Observations and Free-free Emission in High-redshift Star-forming Galaxies}

\author{H. S. B.\ Algera\altaffilmark{1}}
\altaffiltext{1}{Leiden Observatory, Leiden University, P.O. Box 9513, 2300 RA Leiden, the Netherlands}
\email{algera@strw.leidenuniv.nl}

\author{J. A.\ Hodge\altaffilmark{1}}

\author{D.\ Riechers\altaffilmark{2,3}}
\altaffiltext{2}{Department of Astronomy, Cornell University, Ithaca, New York, 14853, USA}
\altaffiltext{3}{Max-Planck Institute for Astronomy, K\"{o}nigstuhl 17, 69117 Heidelberg, Germany}

\author{E.\ J.\ Murphy\altaffilmark{4}}
\altaffiltext{4}{National Radio Astronomy Observatory, 520 Edgemont Road, Charlottesville, VA 22903, USA}

\author{R.\ Pavesi\altaffilmark{2}}

\author{M.\ Aravena\altaffilmark{5}}
\altaffiltext{5}{N\'{u}cleo de Astronom\'{i}a, Facultad de Ingenier\'{i}a y Ciencias, Universidad Diego Portales, Av. Ej\'{e}rcito 441, Santiago, Chile}

\author{E.\ Daddi\altaffilmark{6}}
\altaffiltext{6}{Laboratoire AIM, CEA/DSM-CNRS-Universite Paris Diderot, Irfu/Service dAstrophysique, CEA Saclay, Orme des Merisiers, F-91191, Gif-sur-Yvette cedex, France}

\author{R.\ Decarli\altaffilmark{7}}
\altaffiltext{7}{INAF -- Osservatorio di Astrofisica e Scienza dello Spazio di Bologna, via Gobetti 93/3, I-40129, Bologna, Italy}

\author{M.\ Dickinson\altaffilmark{8}}
\altaffiltext{8}{National Optical Astronomy Observatory, 950 North Cherry Avenue, Tucson, AZ 85719, USA}

\author{M.\ Sargent\altaffilmark{9}}
\altaffiltext{9}{Astronomy Centre, Department of Physics and Astronomy, University of Sussex, Brighton, BN1 9QH, UK}

\author{C.\ E.\ Sharon\altaffilmark{10}}
\altaffiltext{10}{Yale-NUS College, 16 College Avenue West 01-220, 138527 Singapore}

\author{J.\ Wagg\altaffilmark{11}}
\altaffiltext{11}{SKA Organization, Lower Withington Macclesfield, Cheshire SK11 9DL, UK}

\begin{abstract}
The high-frequency radio sky has historically remained largely unexplored due to the typical faintness of sources in this regime, and the modest survey speed compared to observations at lower frequencies. However, high-frequency radio surveys present an invaluable tracer of high-redshift star-formation, as they directly target the faint radio free-free emission. We present deep continuum observations at 34\,GHz in the COSMOS and GOODS-North fields from the \emph{Karl G. Jansky} Very Large Array (VLA), as part of the COLD$z$ survey. The deep COSMOS mosaic spans $\sim10\,\text{arcmin}^2$ down to $\sigma=1.3\,\mu\text{Jy\,beam}^{-1}$, while the wider GOODS-N observations cover $\sim50\,\text{arcmin}^2$ to $\sigma=5.3\,\mu\text{Jy\,beam}^{-1}$. We present the deepest 34\,GHz radio number counts to date, with five and thirteen continuum detections in COSMOS and GOODS-N, respectively. Nine galaxies show 34\,GHz continuum emission that is originating from star-formation, although for two sources this is likely due to thermal emission from dust. Utilizing deep ancillary radio data at 1.4, 3, 5 and 10\,GHz, we decompose the spectra of the remaining seven star-forming galaxies into their synchrotron and thermal free-free components, finding typical thermal fractions and synchrotron spectral indices comparable to those observed in local star-forming galaxies. Using calibrations from the literature, we determine free-free star-formation rates (SFRs), and show that these are in agreement with SFRs from spectral energy distribution fitting and the far-infrared/radio correlation. Our observations place strong direct constraints on the high-frequency radio emission in typical galaxies at high-redshift, and provide some of the first insight in what is set to become a key area of study with future radio facilities as the Square Kilometer Array Phase 1 and next-generation VLA.



\end{abstract}

\section{Introduction}
\label{sec:introduction}
Deep radio observations offer an invaluable view on star-formation in the high-redshift Universe. With current facilities, such as the upgraded NSF's \emph{Karl G. Jansky} Very Large Array (VLA), both star-forming galaxies and faint active galactic nuclei (AGN) can now be studied down to microJansky flux densities at GHz frequencies. However, a regime that remains substantially understudied is the faint radio population at high frequencies ($\nu\gtrsim10$ GHz), which is for a large part the result of the comparative inefficiency at which high-frequency radio surveys can be executed. Firstly, for a fixed telescope size, the field of view of a single pointing decreases steeply with frequency as $\nu^{-2}$. Secondly, radio sources are generally intrinsically fainter at high radio frequencies, $S_\nu \propto \nu^\alpha$, where $\alpha \sim -0.7$ \citep{condon1992}, and thirdly, typically only a modest fraction of telescope observing time is suitable for high-frequency observations due to more stringent requirements on the observing conditions. As a result, the survey speed at 34\,GHz is $\gtrsim5000\times$ smaller compared to observations at the more commonly utilized frequency of 1.4\,GHz.

Despite these observational difficulties, high-frequency radio observations provide complementary insight into both star-forming galaxies and AGN. Historically, low-frequency radio observations have been used as a tracer of star-formation activity through the far-infrared/radio correlation (FIRRC; \citealt{vanderkruit1971,vanderkruit1973,dejong1985,helou1985}). This correlation, which has been shown to hold over several orders of magnitude in terms of luminosity \citep{yun2001,bell2003}, as well as to high-redshift ($z\sim5$, \citealt{delhaize2017,calistrorivera2017,algera2020a,delvecchio2020}), relates the predominantly non-thermal synchrotron emission of a star-forming galaxy to its far-infrared (FIR) luminosity. The latter has been well-calibrated as a tracer of star-formation, as at FIR-wavelengths dust re-emits the light absorbed from young, massive stars (e.g., \citealt{kennicutt1998}). The synchrotron emission, instead, emanates from cosmic rays accelerated by supernova-induced shocks, and as such constitutes a tracer of the end product of massive-star formation \citep{condon1992,bressan2002}. However, a second process is expected to dominate the radio spectral energy distribution (SED) at high frequencies ($\nu\gtrsim30\,$GHz): radio free-free emission (FFE). Unlike radio synchrotron radiation, free-free emission is a much more direct star-formation rate tracer, as it originates from the HII-regions in which massive stars have recently formed. In addition, unlike other commonly used probes of star-formation such as UV continuum emission or the hydrogen Balmer lines, free-free emission constitutes a tracer of star-formation that is largely unbiased by dust extinction. These characteristics establish free-free emission as one of the most reliable tracers of star formation, both in the local and high-redshift Universe.

Locally, the radio spectra of both individual star-forming regions and star-forming galaxies have been well-characterized, and have established free-free emission as a means of calibrating other tracers of star-formation \citep{murphy2011}. In addition, in nearby galaxies free-free emission can typically be separated spatially from non-thermal synchrotron emission, as individual extragalactic star-forming regions can be resolved \citep{tabatabaei2013,querejeta2019,linden2020}. However, in the high-redshift Universe, free-free emission has remained elusive, despite the observational advantage that high-frequency continuum emission redshifts into radio bands that are more easily accessible from Earth, facilitating the sampling of the free-free-dominated regime of the radio spectrum. The comparative faintness of high-redshift galaxies, however, complicates the usage of free-free emission as a tracer of star-formation at early cosmic epochs. Indeed, current detections of high-frequency continuum emission in distant galaxies remain limited to bright or gravitationally lensed starbursts \citep{thomson2012,aravena2013,riechers2013,wagg2014,huynh2017,penney2020}. In addition, most of these studies lacked the ancillary low-frequency data required to robustly disentangle free-free emission from the overall radio continuum, which requires observations at at least three frequencies (e.g., \citealt{tabatabaei2017,klein2018}), or probed rest-frame frequencies dominated by thermal emission from dust ($\nu\gtrsim200\,$GHz; \citealt{condon1992}). Despite this observational complexity, one of the key science goals for upcoming radio facilities such as the next-generation VLA is to systematically use free-free emission as a probe of star-formation in the high-redshift galaxy population \citep{barger2018}. As such, it is of considerable interest to already explore this high-frequency parameter space with current radio facilities.

While radio continuum observations are invaluable in characterizing high-redshift star-formation in a dust-unbiased manner, radio surveys are additionally capable of detecting what fuels this process, namely molecular gas. For a clear census of the molecular gas reservoir of the Universe, blind surveys are crucial, as they do not suffer from any biases arising from follow-up radio observations of known high-redshift sources. The first such blind surveys have recently been completed, such as ASPECS \citep{walter2016,decarli2016}, and the CO Luminosity Density at High Redshift survey (COLD$z$; \citealt{pavesi2018,riechers2019,riechers2020}), which targets low-$J$ CO observations at 34\,GHz using the VLA. Due to the large bandwidth of its high-frequency receivers, deep VLA surveys of molecular gas result in sensitive continuum images essentially `for free'. In this work, we describe the deep continuum observations of the COLD$z$ survey. Our main goal is to constrain the radio spectra of typical sources in a frequency range that has not been widely explored, and extend this to a new parameter space of faint AGN and star-forming galaxies, down to the $\mu$Jy level. The COLD$z$ survey covers a region of two well-studied extragalactic fields, COSMOS \citep{scoville2007} and GOODS-North \citep{giavalisco2004}, and hence allows for a multi-wavelength perspective on this faint population.

The outline of this paper is as follows. In Section \ref{sec:observations} we describe our 34\,GHz VLA observations and the creation of deep continuum images, as well as the existing ancillary multi-wavelength data. We outline the detection and source extraction of sources at 34\,GHz in Section \ref{sec:sourcedetection}, and assign the radio sample multi-wavelength counterparts and redshifts in Section \ref{sec:multiwavelength}. We present our deep radio number counts at 34\,GHz in Section \ref{sec:sourcecounts}, and separate the sample into radio AGN and star-forming galaxies in Section \ref{sec:radioproperties}. In this Section, we additionally decompose the spectra of the star-forming galaxies into radio synchrotron and free-free emission, and compare the latter as a tracer of star-formation with more commonly adopted tracers at high-redshift. Section \ref{sec:nextgen} provides an outlook for the future, and discusses how upcoming radio facilities will revolutionize high-redshift studies of radio star-formation. Finally, we summarize our findings in Section \ref{sec:conclusion}. Where necessary, we assume a standard $\Lambda$-Cold Dark Matter cosmology, with $H_0=70\,\text{km\,s}^{-1}\text{\,Mpc}^{-1}$, $\Omega_m=0.30$ and $\Omega_\Lambda=0.70$. Magnitudes are quoted in the AB-system and a \citet{chabrier2003} initial mass function is assumed.

\section{OBSERVATIONS \& DATA REDUCTION}
\label{sec:observations}

\subsection{COLD$z$}
The COLD$z$ survey \citep{pavesi2018,riechers2019,riechers2020} was designed to blindly probe the low$-J$ CO-lines in high-redshift galaxies, which requires high-frequency observations spanning a large bandwidth. Both are provided by the Ka-band of the upgraded VLA, which allows for a total of 8\,GHz of contiguous frequency coverage, tuneable to be within the range of $26.5-40.0\,$GHz. The COLD$z$ observing strategy is presented in \citet{pavesi2018}, to which we refer the reader for additional details.

For the optimal constraints on the CO luminosity function, COLD$z$ combines a deep yet small mosaic in the COSMOS field with wider but shallower observations in the GOODS-N field. The COSMOS observations constitute a 7-pointing mosaic, and account for a total on-source time of 93\,hr. The data were taken in the VLA D (82\,hr) and DnC (11\,hr) configurations, and span a total frequency range of $30.969 - 39.033\,$GHz, using dual polarization. The observations of the GOODS-N field comprise a total of 122\,hr of on-source time, and make up a 57-pointing mosaic. The data were predominantly taken in the D-configuration (83.5\,hr), with additional observations taking place in the D$\rightarrow$DnC (4.4\,hr), DnC (30.8\,hr) and DnC$\rightarrow$C (3\,hr) configurations. The observations cover a frequency range between $29.981 - 38.001\,$GHz, and in what follows we will refer to both the COSMOS and GOODS-N observations by their typical central frequency of 34\,GHz.

Calibration of the observations was done in {\sc{CASA}} version 4.1.0, with extensive use being made of a modified version of the VLA pipeline, the details of which can be found in \citet{pavesi2018}. As one of the main goals of COLD$z$ is to detect spectral lines, both Hanning smoothing and {\sc{RFLAG}}, used to remove radio frequency interference (RFI), were switched off. Instead, some persistent RFI at 31.5\,GHz was flagged manually, and some occasions of narrow noise spikes were flagged via the methods detailed in \citet{pavesi2018}. The data were then concatenated, and, for the continuum observations presented here, subsequently averaged in time (9\,s) and frequency (16\,channels), in order to reduce the size of the dataset prior to imaging. The imaging of the calibrated 34\,GHz observations was carried out in CASA version 4.3.1, using the `mosaic' mode of CASA task {\sc{clean}}. For both the COSMOS and the GOODS-N mosaics, a multi-frequency synthesis algorithm was employed to take into account the large bandwidth of the observations. A natural weighting was further adopted, in order to maximize the sensitivity of the data. The data were imaged iteratively, by cleaning all sources at $>6\sigma$ down to the $2\sigma$ level. \\

\begin{figure*}
    \centering
    \includegraphics[width=0.556\textwidth]{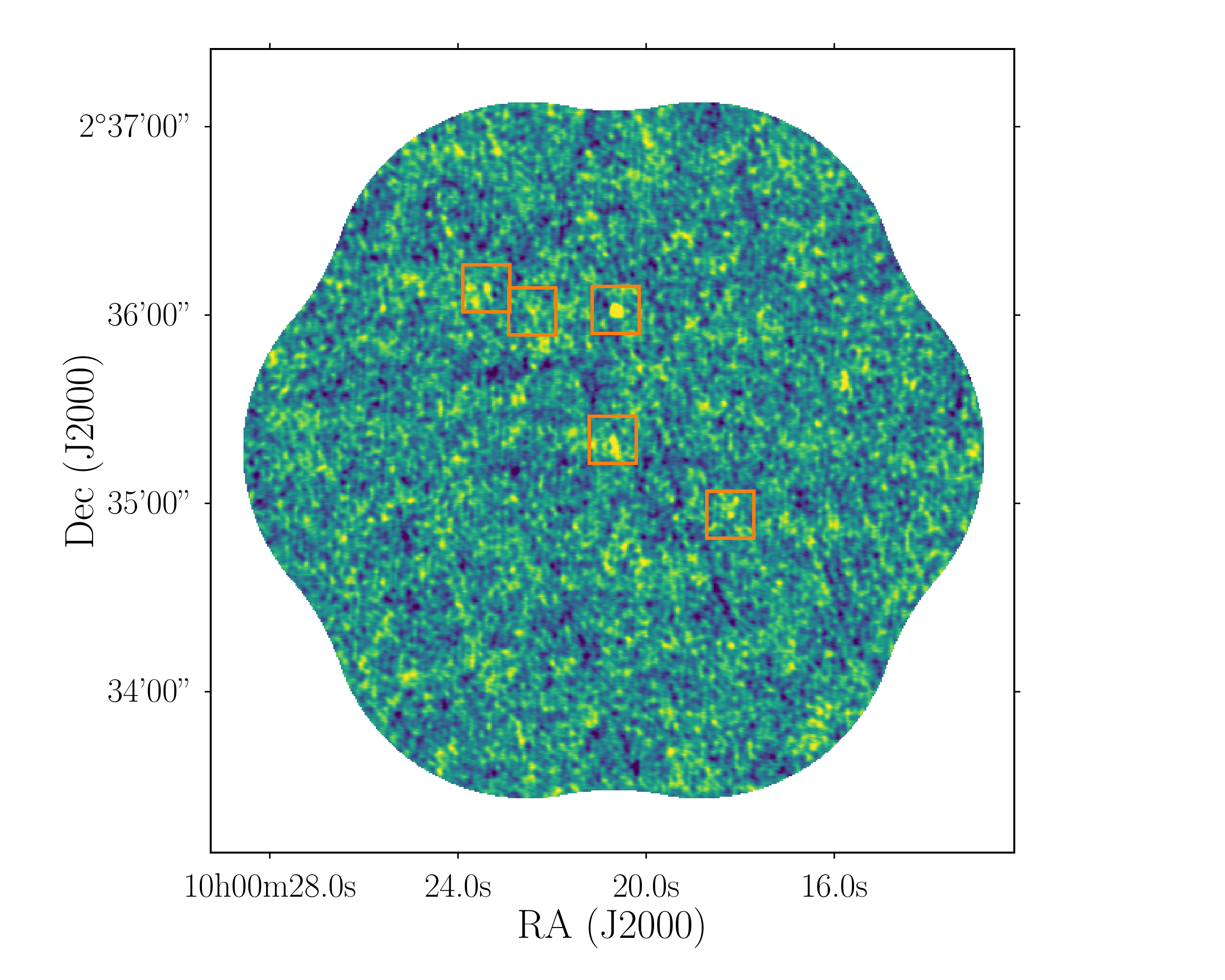} \hspace*{-0.5cm}
    \includegraphics[width=0.444\textwidth]{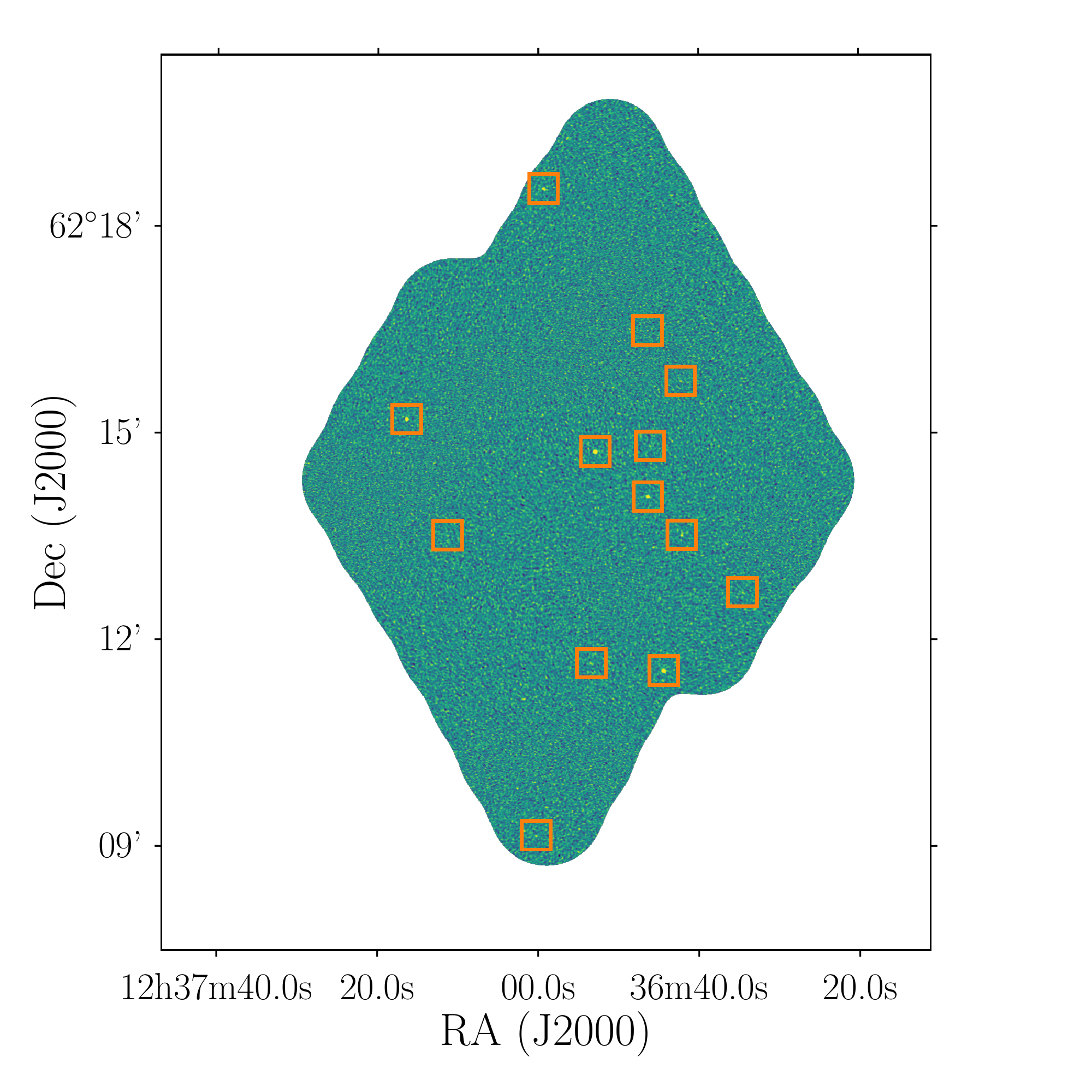}
    \caption{\textbf{Left:} The 7-pointing COSMOS 34\,GHz continuum mosaic of $\sim10\,\text{arcmin}^2$. The five robust detections (Section \ref{sec:sourcedetection}) are highlighted via orange squares. \textbf{Right:} The $\sim 50\,\text{arcmin}^2$ GOODS-North 34\,GHz mosaic (pointings $1-56$), with the thirteen robust detections highlighted. Both mosaics are uncorrected for primary beam attenuation, and the colorscale runs from $-3\sigma$ to $3\sigma$, where $\sigma$ represents the (uncorrected) RMS-noise in the image. It is clear that many 34-GHz detections lie close to the $3\sigma$ detection threshold, and would not have been identified without deep, ancillary radio data across the fields.}
    \label{fig:mosaics}
\end{figure*}

\begin{figure*}
    \centering
    \includegraphics[width=0.556\textwidth]{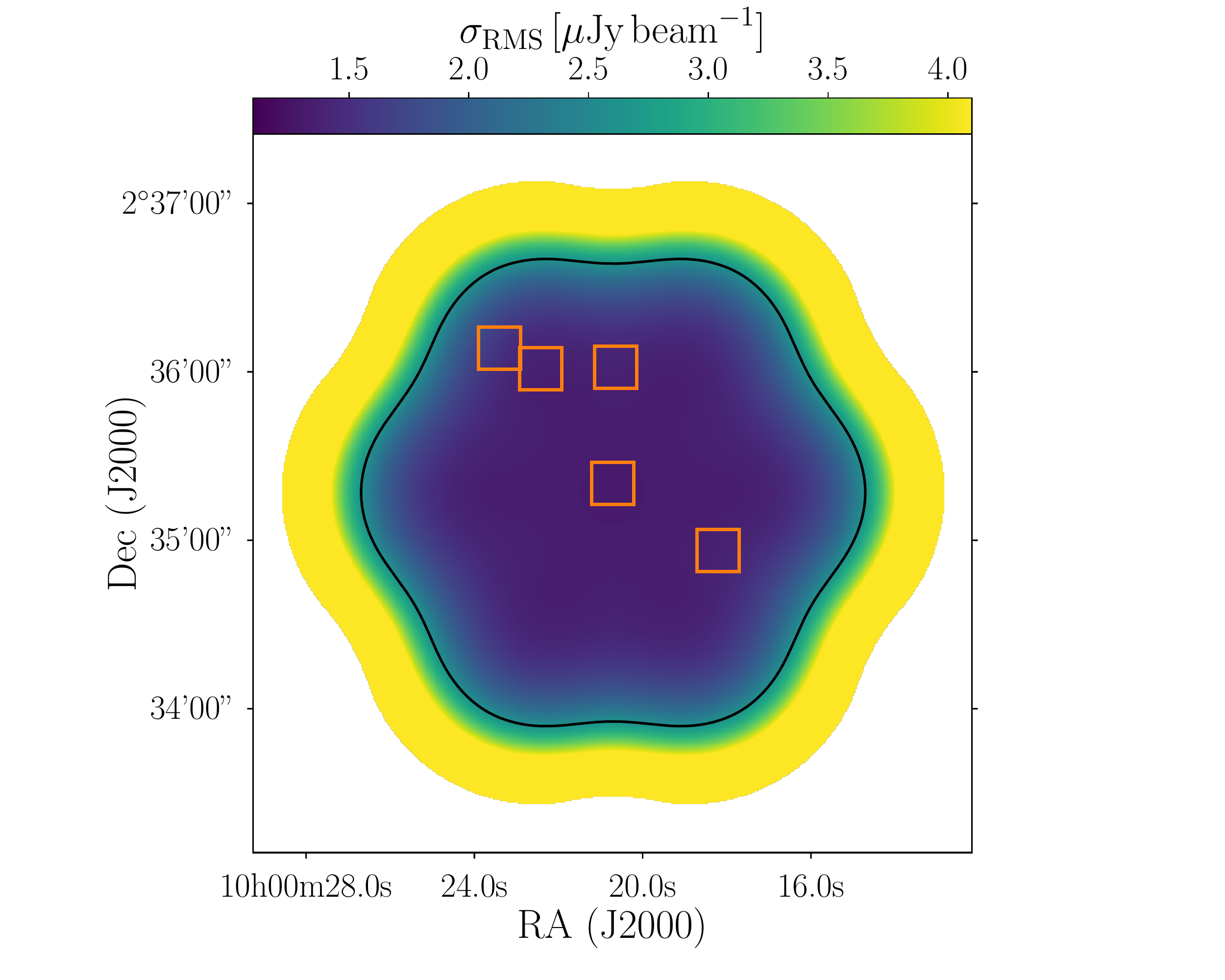} \hspace*{-0.5cm}
    \includegraphics[width=0.444\textwidth]{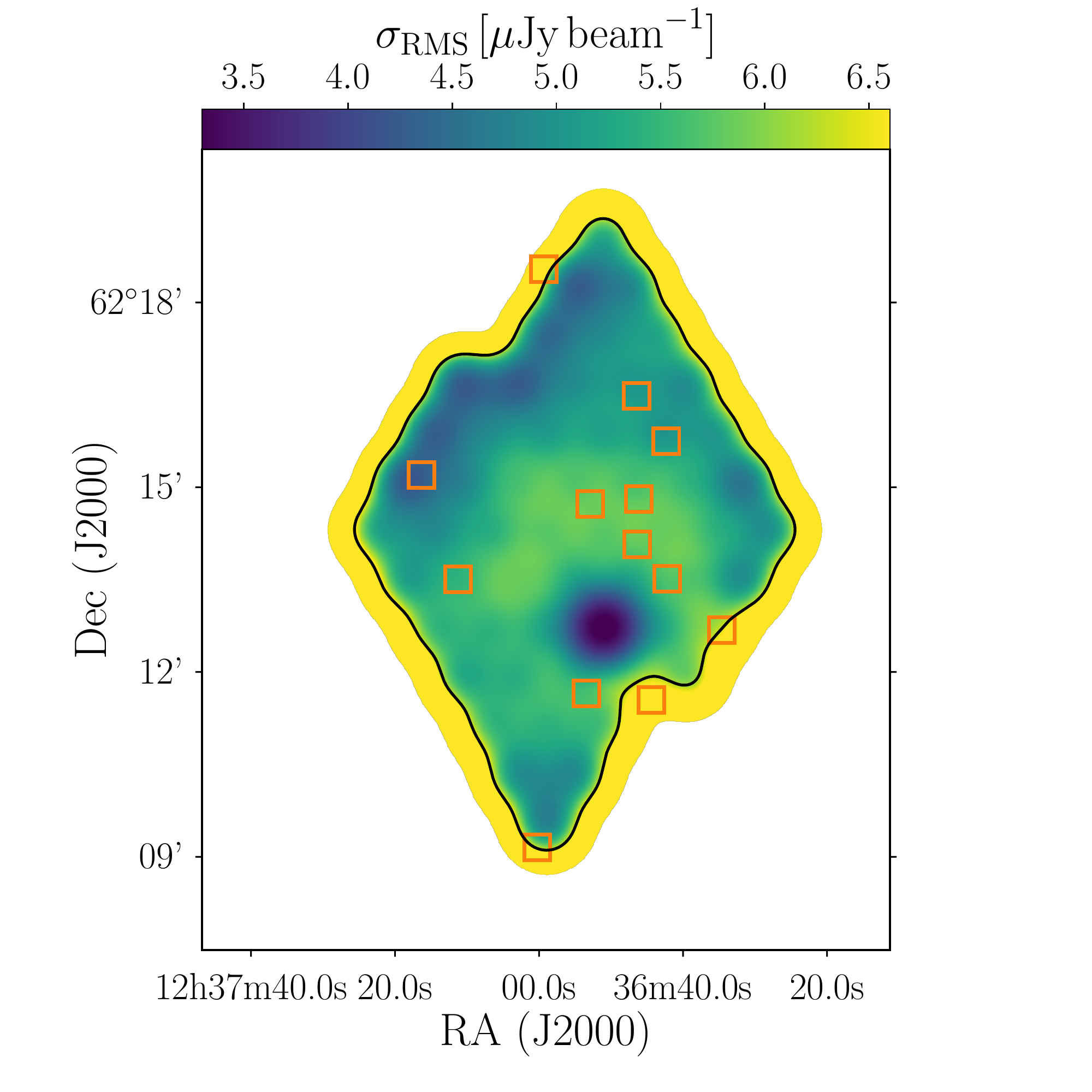}
    \caption{\textbf{Left:} The COLD$z$ COSMOS RMS-map, after correction for primary beam attenuation. The uncorrected RMS-map is highly uniform, and any structure in the map is solely the result of the beam correction. \textbf{Right:} The RMS-map of the 34-GHz observations across the GOODS-N field. The map uncorrected for the beam is highly uniform in its noise properties, and the structure in the map indicates the differences in observing time across the mosaic, with a single deep pointing centered on $\text{R.A.}=\,$12:36:51.06, $\text{decl.}=\,$+62:12:43.8, designed to overlap with the PdBI observations from \citet{decarli2014}, highlighted as the region with the lowest RMS. The black lines outlining the mosaics indicate the region where the primary beam reaches 50\% of its maximum sensitivity, and the robust detections in both fields are highlighted via the orange boxes as in Figure \ref{fig:mosaics}.}
    \label{fig:rmsmaps}
\end{figure*}

We present the 34\,GHz continuum maps across the COSMOS and GOODS-N fields, as well as the corresponding RMS-maps, in Figures \ref{fig:mosaics} and \ref{fig:rmsmaps}. The COSMOS mosaic covers a field-of-view of $9.6\,\text{arcmin}^2$ out to $20\%$ of the peak primary beam sensitivity. The central RMS-noise in the image equals $1.3\,\mu\text{Jy\,beam}^{-1}$, with the typical RMS increasing to $1.5\,\mu\text{Jy\,beam}^{-1}$ and $1.9\,\mu\text{Jy\,beam}^{-1}$, within 50 and 20 per cent of the peak primary beam sensitivity, respectively. The synthesized beam of the COSMOS observations is well-described by an elliptical Gaussian of $2\farcs70\times2\farcs41$ with a position angle of $-0.7^\circ$.

The GOODS-N mosaic spans an area of $51\,\text{arcmin}^2$ out to 20\% of the peak primary beam sensitivity, covering roughly a third of the `traditional' GOODS-N survey \citep{giavalisco2004}. The typical noise level varies slightly across the mosaic, with a single, deep 34\,GHz pointing in the field reaching a central noise level of $3.2\,\mu\text{Jy\,beam}^{-1}$. Across the entire GOODS-N field, within 50 and 20 per cent of the peak primary beam sensitivity, respectively, the typical RMS-noise equals $5.3\,\mu\text{Jy\,beam}^{-1}$ and $5.5\,\mu\text{Jy\,beam}^{-1}$. The synthesized beam is well-described by an elliptical Gaussian of $2\farcs19\times1\farcs84$, with a position angle of $75.3^\circ$, after smoothing the different pointings to a common beam. \citet{pavesi2018} also provide a version of the GOODS-N radio map where all pointings have been imaged at their native resolution. While the resulting mosaic cannot be described by a single beam, it allows for the search for unresolved sources at slightly higher signal-to-noise, as no smoothing or tapering was required. We use this ``unsmoothed'' map for source detection in Section \ref{sec:sourcedetection}, in addition to the regular mosaic with a homogenized beam.

\subsection{Ancillary Radio Observations}
The COLD$z$ observations in the COSMOS field overlap in their entirety with deep 3 and 10\,GHz VLA pointings from the COSMOS-XS survey \citep{vandervlugt2020,algera2020b}. These images have a resolution similar to COLD$z$, with a synthesized beam of $2\farcs14\times1\farcs81"$ and $2\farcs33\times 2\farcs01$ at 3 and 10\,GHz, respectively. At 3\,GHz, the observations reach a central RMS of $0.53\,\mu\text{Jy\,beam}^{-1}$, with a typical primary beam sensitivity of $90\%$ of the maximum across the COLD$z$ mosaic. The 10\,GHz data reach a central RMS-sensitivity of $0.41\,\mu\text{Jy\,beam}^{-1}$ (typical primary beam sensitivity of $80\%$), and were centered on the COLD$z$ observations by design. In total, 70 sources detected at 3\,GHz at $\geq5\sigma$ fall within the $\sim10\,\text{arcmin}^2$ COLD$z$ field-of-view. A subset of 40 are additionally detected at 10\,GHz. In addition, the central region of the COSMOS field, spanning $\approx1\,\text{deg}^2$, is covered by VLA observations at 1.4\,GHz \citep{schinnerer2007,schinnerer2010}. These observations reach a typical RMS sensitivity of $12\,\mu\text{Jy\,beam}^{-1}$ across the COLD$z$ field-of-view, at a resolution of $1\farcs5\times 1\farcs4$. In total, two sources within the COLD$z$ field-of-view are detected at 1.4\,GHz. Adopting a typical spectral index of $\alpha = -0.70$, the COLD$z$ and 1.4\,GHz COSMOS observations are approximately of equal relative depth.\footnote{However, we note that we adopt a $3\sigma$ detection threshold for the COLD$z$ data in Section \ref{sec:sourcedetection}, whereas a $5\sigma$ threshold was adopted by \citet{schinnerer2007} for the 1.4\,GHz observations. As such, sources detected at $3\sigma$ in the COLD$z$ survey require a spectral index of $\alpha \lesssim -0.85$ to additionally be detected at 1.4\,GHz.}

The GOODS-N field is similarly well-covered by radio observations from the VLA. At 1.4\,GHz, the entire field has been imaged in a single pointing by \citet{owen2018}, down to a central RMS of $2.2\,\mu\text{Jy\,beam}^{-1}$ with little variation across the field-of-view of the COLD$z$ mosaic. The resolution of the radio data equals $1\farcs6$, and a total of 186 sources detected at 1.4\,GHz by \citet{owen2018} fall within the 20 percent power point of the GOODS-N COLD$z$ image. The field has additionally been mapped at 5\,GHz by \citet{gim2019}. They detect 52 sources down to an RMS of $3.5\,\mu\text{Jy\,beam}^{-1}$ across two VLA pointings covering a total area of $109\,\text{arcmin}^2$, which fully overlaps with the COLD$z$ footprint. Their angular resolution of $1\farcs47\times1\farcs42$ is similar to that of the deeper 1.4\,GHz observations, and \citet{gim2019} find that all 5\,GHz detections have a counterpart in this lower frequency map. Finally, \citet{murphy2017} present a single pointing at 10\,GHz in the GOODS-N field (primary beam FWHM of $4\farcm25$) at a native resolution of $0\farcs22$. At this high angular resolution, their observations reach a sensitivity of $0.57\,\mu\text{Jy\,beam}^{-1}$ in the center of the pointing. \citet{murphy2017} further provide two tapered images with a resolution of $1''$ and $2''$, similar to the ancillary radio data in the field. These images reach an RMS-noise of $1.1$ and $1.5\,\mu\text{Jy\,beam}^{-1}$, respectively. In total, \citet{murphy2017} recover 38 sources across the combined high-resolution and tapered images. Approximately $75\%$ of the COLD$z$ GOODS-N data overlap with the smaller pointing at 10\,GHz, when imaged out to 5\% of the primary beam FWHM \citep{murphy2017}. 

\begin{figure}
    \centering
    \hspace*{-0.2cm}
    \includegraphics[width=0.5\textwidth]{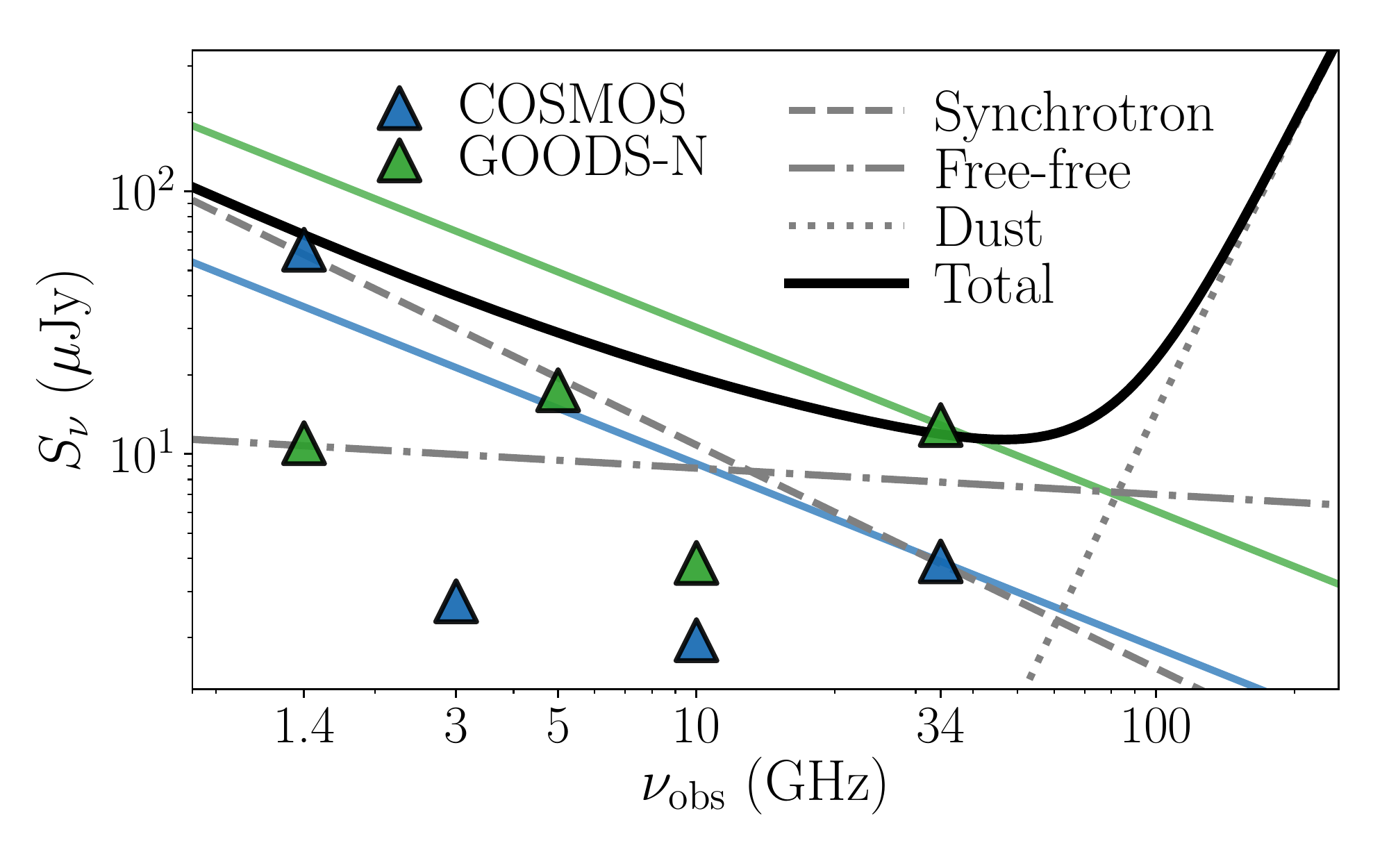}
    \caption{The detection limits of the various radio observations across COSMOS and GOODS-N, superimposed on the radio spectrum of a star-forming galaxy ($\text{SFR}=100\,M_\odot\,\text{yr}^{-1}$ at $z=1$; see text for details). The blue (green) lines indicate the detection limit of the COLD$z$ 34\,GHz continuum data, scaled with a typical $\alpha=-0.70$. The different constituents of the radio spectrum are shown, including free-free emission which is expected to become dominant at rest-frame $\nu\gtrsim30\,$GHz. As such, the COLD$z$ observations directly target the strongly free-free-dominated regime.}
    \label{fig:example}
\end{figure}

We summarize the detection limits of the various radio observations across both COSMOS and GOODS-N in Figure \ref{fig:example}, and compare these with the typical radio spectrum of a star-forming galaxy ($\text{SFR}=100\,M_\odot\,\text{yr}^{-1}$ at $z=1$). This assumes the \citet{kennicutt1998} conversion between SFR and infrared luminosity, adapted for a Chabrier IMF, as well as a simple optically thin dust spectral energy distribution with $\beta = 1.8$ and $T_\text{dust} = 35\,$K. We further adopt the \citet{condon1992} model for the radio spectrum of star-forming galaxies, and assume the far-infrared/radio correlation from \citet{delhaize2017}. Under these assumptions, a galaxy with $\text{SFR} = 100\,M_\odot\,\text{yr}^{-1}$ can be directly detected at 34\,GHz out to $z=2$ ($z=1$) in the COSMOS (GOODS-N) field.

\subsection{Ultra-violet to Submillimeter Observations}

Both the COSMOS and GOODS-North fields have been targeted by a wealth of multi-wavelength observations, spanning the full X-ray to radio regime. For the COSMOS field, we adopt the multi-wavelength matching procedure for the COSMOS-XS survey \citep{algera2020b}, as all 34\,GHz continuum detections have radio counterparts in this survey (Section \ref{sec:sourcedetection}). This cross-matching procedure invokes the recent ``Super-deblended'' catalog from \citet{jin2018}, who adopt a novel deblending technique to address confused mid-infrared to submillimeter observations. The Super-deblended catalog provides photometry from \emph{Spitzer}/IRAC $3.6\,\mu$m to MAMBO $1.2\,$mm (for a full list of references see \citealt{jin2018}), as well as radio data at 1.4 and 3\,GHz from \citet{schinnerer2010} and \citet{smolcic2017a}, respectively. However, we directly adopt the radio fluxes from the 1.4\,GHz catalog from \citet{schinnerer2007,schinnerer2010}, and use the deeper COSMOS-XS 3\,GHz data in favor of the observations from \citet{smolcic2017a}. In addition, \citet{algera2020b} cross-match with the $z^{++}YJHK_s$-selected COSMOS2015 catalog from \citet{laigle2016} containing far-ultraviolet to near-infrared photometry, in order to complete the coverage of the spectral energy distribution. Finally, we search for Atacama Large Millimeter/submillimeter Array (ALMA) counterparts as part of the AS2COSMOS \citep{simpson2020} and A3COSMOS surveys \citep{liu2019}, which constitute a collection of individual pointings across the COSMOS field.

A wealth of multi-wavelength data similarly exist across the GOODS-N field. In order to obtain optical/near-infrared photometry for the 34\,GHz continuum detections, we adopt the 3D-HST photometric catalog from \citet{skelton2014}. Source detection was performed in a deep combined F125W+F140W+F160W image, and further photometry is carried out in the wavelength range of $0.3-8\,\mu$m, including \emph{Spitzer}/IRAC observations from \citet{dickinson2003} and \citet{ashby2013}. For further details, we refer the reader to \citet{skelton2014}. We obtain additional mid- and far-infrared photometry from the Super-deblended catalog across the GOODS-N field by \citet{liu2018}. They adopt combined priors from \emph{Spitzer}/IRAC, \emph{Spitzer}/MIPS $24\,\mu$m and VLA 1.4\,GHz observations, and utilize these to deblend the photometry at more strongly confused wavelengths. The Super-deblended catalog provides additional photometry from \emph{Spitzer}/IRS, \emph{Herschel}/PACS \citep{magnelli2013} and \emph{Herschel}/SPIRE \citep{elbaz2011}, as well as data from JCMT/SCUBA-2 $850\,\mu$m \citep{geach2017} and AzTEC+MAMBO 1.2\,mm \citep{penner2011}.

\subsection{X-ray Observations}
While a radio-based selection renders one sensitive to AGN activity at radio wavelengths, $\sim20-25\%$ of the faint radio population ($S_{1.4} \lesssim 1\,$mJy; equivalent to $S_{34} \lesssim 100\,\mu$Jy given $\alpha=-0.70$) is thought to consist of radio-quiet AGN \citep{bonzini2013,smolcic2017b}. These sources show no substantial AGN-related emission at radio wavelengths, but are instead classified as AGN based on signatures at other wavelengths. Strong X-ray emission, in particular, forms an unmistakable manifestation of AGN activity. As such, we make use of the deep X-ray coverage over both the COSMOS and GOODS-N fields to characterize the nature of the 34\,GHz continuum detections.

The COSMOS field is covered in its entirety by the 4.6 Ms \emph{Chandra} COSMOS Legacy survey \citep{civano2016}, with the individual \emph{Chandra} pointings accounting for a typical $\approx160\,$ks of exposure time. The area covered by the COLD$z$ survey contains $3$ X-ray detections, at a typical detection limit of the survey of $\sim2\times10^{-15}\,\text{erg\,cm}^{-2}\text{s}^{-1}$ in the full range of $2-10\,$keV. The catalog provided by \citet{marchesi2016} further includes X-ray luminosities for X-ray detections with robust optical and infrared counterparts.

The GOODS-North field is similarly covered by deep 2 MS \emph{Chandra} observations as part of the \emph{Chandra} Deep Field North Survey \citep{xue2016}. Across the COLD$z$ field-of-view, the survey identifies 189 X-ray sources, and attains a flux limit of $\sim3.5\times10^{-17}\,\text{erg\,cm}^{-2}\,\text{s}^{-1}$ in the $0.5-7\,$keV energy range -- nearly a factor of $50$ deeper than the COSMOS data, when converted to the same energy range adopting $\Gamma = 1.8$. The catalog provided by \citet{xue2016} further includes X-ray luminosities for the entries with reliable redshift information.

\section{CONTINUUM SOURCES}
\label{sec:sourcedetection}

\begin{figure*}[t]
    \centering
    \includegraphics[width=\textwidth]{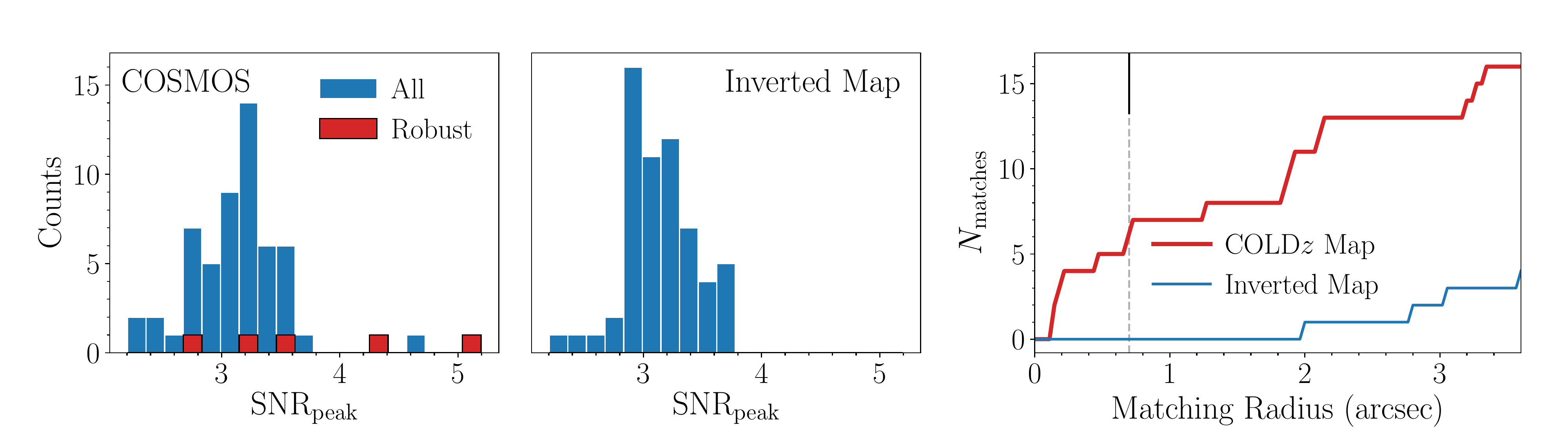}
    \caption{\textbf{Left:} distribution of signal-to-noise ratios for all the peaks identified in the COSMOS 34\,GHz image. Sources with lower-frequency radio counterparts within $0\farcs7$ are highlighted in red. The most significant detection at $\text{SNR}_{34} = 17$ is not shown for clarity. \textbf{Middle:} same as the left panel, now showing ``sources'' detected in the inverted image. No matches within $0\farcs7$ are found with lower-frequency radio counterparts, and no spurious detection above $\text{SNR}=3.7$ exists. \textbf{Right:} number of cross-matches between the 34\,GHz continuum detections and the detections in the deeper 3\,GHz COSMOS-XS map, as a function of matching radius. The radius adopted in this work is indicated via the black vertical line. No spurious matches are found in the inverted map out to $\sim2''$, indicating the sources recovered in the 34\,GHz map are likely to be real.}
    \label{fig:COSMOS_SNR}
\end{figure*}

\begin{figure*}[t]
    \centering
    \includegraphics[width=\textwidth]{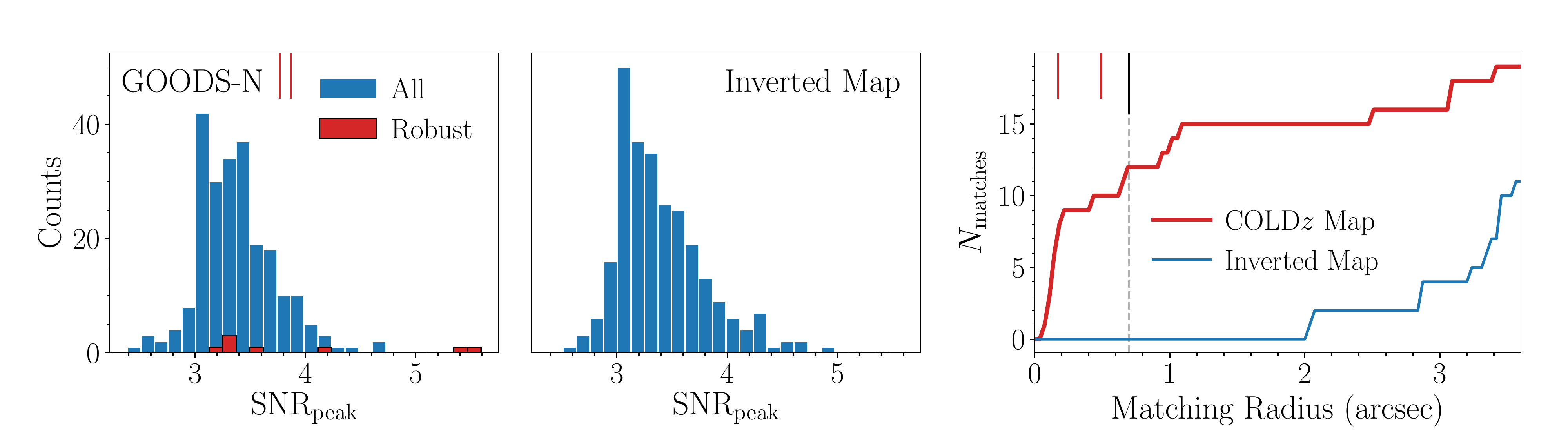}
    \caption{Same as Figure \ref{fig:COSMOS_SNR}, now showing peaks identified in the GOODS-N radio map. Four high-SNR detections in the real map are not shown for clarity, and the two red, vertical dashes indicate the two additional sources we recover when performing source detection in the unsmoothed radio map (see text for details). In total, 13 sources are robustly detected at 34\,GHz in GOODS-N.}
    \label{fig:GOODSN_SNR}
\end{figure*}

\begin{figure*}[!t]
\includegraphics[width=\textwidth]{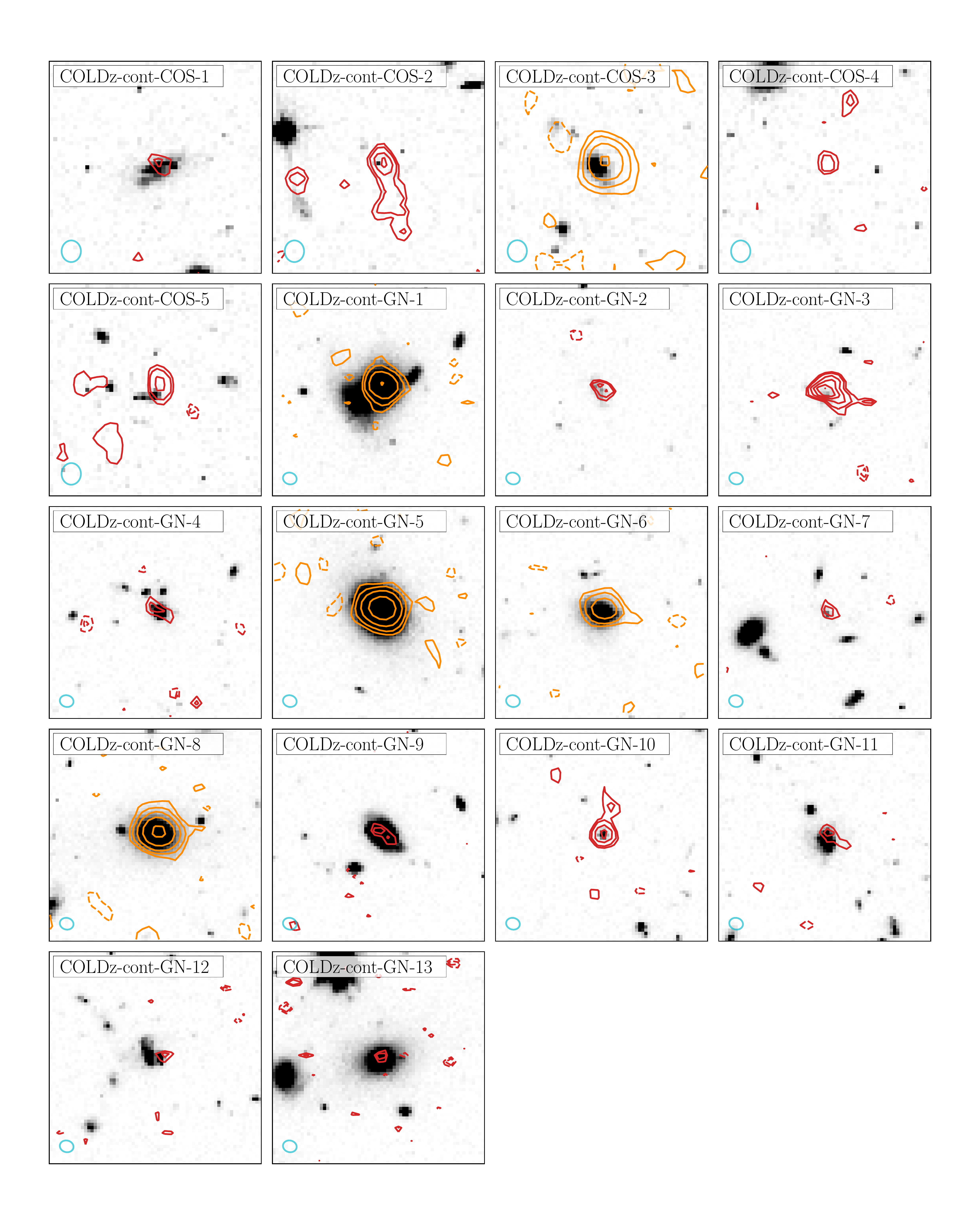}
    \caption{Postage stamps ($16''\times16''$) of the 18 robust COLD$z$ detections, overlaid on \emph{HST} F814W (COSMOS) or F160W (GOODS-N) images. For faint sources ($\text{SNR}_{34} \leq 10$), 34\,GHz contours are shown in red, and represent $\pm2.5\sigma$, followed by $\pm3\sigma,\pm4\sigma\ldots$ in steps of $1\sigma$, where $\sigma$ represents the local RMS in the radio map. Bright sources -- typically radio AGN, as will be quantified in Section \ref{sec:radioagn} -- are shown via orange contours, in steps of $\pm2^N\sigma$, where $N=1,2,\ldots$. Negative contours are shown via dashed lines, and the beam size is indicated in the lower left corner via the cyan ellipse. While half of the sample is detected at relatively low signal-to-noise ($3-5\sigma$), most of these are star-forming galaxies, likely dominated at these frequencies by radio free-free emission (Section \ref{sec:decomposition}), and form the focus of this work.}
    \label{fig:sources}
\end{figure*}

We run source detection on our 34\,GHz radio maps using {\sc{PyBDSF}} \citep{mohanrafferty2015}, prior to correcting the images for the primary beam. This has the benefit that the noise properties are uniform across the mosaics, which facilitates source detection, and additionally ensures that fewer spurious sources arise around the noisy edges of the maps.\footnote{We have verified that, after applying the primary beam correction, the flux densities are consistent with those obtained from running source detection on the primary-beam-corrected map, with a typical ratio of $S_\text{uncorr} / S_\text{corr} = 1.05 \pm 0.07$.} In our source detection procedure we can afford to set a liberal detection threshold, as both the COSMOS and GOODS-N fields contain additional low-frequency radio data of greater depth relative to the COLD$z$ observations, and as such we expect any 34\,GHz detections to have radio counterparts. In this work, we therefore adopt a $3\sigma$ peak detection threshold for both images. 

In the COSMOS field, we compare to the deep S- and X-band observations from \citet{vandervlugt2020}. For a source to be undetected in this 10\,GHz map, yet detected at 34\,GHz, requires a highly inverted spectral index of $\alpha^{10}_{34} \gtrsim 0.55$. In the GOODS-N field, we compare with the 1.4\,GHz observations from \citet{owen2018}, and find that 34\,GHz sources require an inverted spectral index of $\alpha^{1.4}_{34} \gtrsim 0.15$ to remain undetected in the lower-frequency map. We note that, as we perform source detection at low signal-to-noise, the source sizes calculated via {\sc{PyBDSF}} may be affected by the local noise properties within the images. In particular, sources with fitted radio sizes smaller than the synthesized beam will have an integrated flux density smaller than the peak value, and will have an uncertainty on the latter that is smaller than the local RMS-noise in the map \citep{condon1997}. We show in Appendix \ref{app:bondi} that all but one source are likely to be unresolved at $\sim2\farcs5$ resolution. For all unresolved sources, we adopt the peak brightness, while the integrated flux density is used otherwise. Following the discussion above, we redefine the uncertainty on the peak brightness and conservatively adopt the maximum of the calculated uncertainty from {\sc{PyBDSF}} and the local RMS at the source position. \\

In the COSMOS map we detect 57 peaks at $\geq3\sigma$ within 20\% of the peak primary beam sensitivity (Figure \ref{fig:COSMOS_SNR}). We can match six to COSMOS-XS counterparts within $0\farcs7$, where we expect to have $N_\text{false} = 0.2$ false matches based on randomly shifting the coordinates of all $\geq3\sigma$ peaks in the mosaic, and repeating the matching a large number of times. This radius was chosen to include all close associations to COSMOS-XS sources, while minimizing the number of expected false matches ($N_\text{false} \ll 1)$. To further verify the robustness of the six close counterparts, we run our source detection procedure on the inverted radio map (i.e., multiplied by $\times-1$), and find a total of 60 negative ``sources'', all of which are by definition spurious. None of these can be matched to COSMOS-XS counterparts within $0\farcs7$, indicating the real matches are likely to be robust. However, out of the six associations to COSMOS-XS galaxies within $0\farcs7$, we find that one candidate source ($\text{SNR}_{34} = 3.3$) is detected solely at 3\,GHz while a 10\,GHz counterpart is also expected, implying it is likely to be spurious. As such, we discard it from our sample and retain 5 sources which form the robust COSMOS continuum sample. Three of these have a signal-to-noise ratio larger than the highest peak-SNR in the inverted radio map of $3.7\sigma$. The remaining two have a relatively low SNR of $\sim3.5$, but are deemed robust due to their multi-wavelength associations. \\

We adopt an identical source detection procedure for the GOODS-N 34\,GHz map. In total, {\sc{PyBDSF}} identifies $236$ peaks above $3\sigma$ in the map, of which we match 12 with low-frequency radio counterparts in the catalog from \citet{owen2018} at $0\farcs7$, where we expect $N_\text{false} = 0.3$ incorrect identifications. We find $263$ sources in the inverted radio map, none of which are matched to lower-frequency radio counterparts within $0\farcs7$. However, upon cross-matching our 12 robust sources with the 5\,GHz catalog from \citet{gim2019}, we find only 11 matches within $0\farcs7$. The single unmatched source ($\text{SNR}_{34} = 3.3$) also falls within field-of-view of the single deep X-band pointing from \citet{murphy2017}, but is additionally undetected at 10\,GHz. As such, this source is likely to be spurious, and we discard it from further analysis. The signal-to-noise distributions of all peaks identified by {\sc{PyBDSF}} in the GOODS-N image and its inverted counterpart are shown in Figure \ref{fig:GOODSN_SNR}. The maximum SNR in the inverted map is $\text{SNR}\approx5.0$, which implies that 5/11 robust 34\,GHz detections lie below the most significant spurious source. This emphasizes the added value of the deep, low-frequency data, which allow us to identify faint sources that would not have been recovered in a blind source detection procedure. We further perform source detection via {\sc{PyBDSF}} on the unsmoothed mosaic (Section \ref{sec:observations}), using the same detection threshold as adopted for the regular GOODS-N mosaic, in order to maximize the number of recovered sources. We find two additional matches with low-frequency counterparts at both 1.4 and 5\,GHz, resulting in a total of 13 sources identified in GOODS-N. We adopt the peak brightness for these two sources, given the lack of a common beam in the unsmoothed mosaic, and use the local RMS in the map as the appropriate uncertainty.

Given that we require any identifications at 34\,GHz to have robust low-frequency radio counterparts, we may be missing any sources with highly unusual inverted spectra. To investigate this possibility, we median-stack in the low-frequency radio maps on the positions of the highest signal-to-noise peaks at 34\,GHz for which no radio counterpart was found (3 and 10\,GHz in COSMOS, 1.4\,GHz in GOODS-N, using the publicly available radio map from \citealt{morrison2010}). In neither the COSMOS nor GOODS-N fields do we see any evidence for positive signal in the stacks, indicating that most -- and likely all -- of these $\text{SNR}\approx3-5$ peaks at 34\,GHz are spurious. In addition, we cross-match the peaks in the 34\,GHz map that do not have radio counterparts with optical/NIR-selected sources from the COSMOS2015 and 3D-HST catalogs in COSMOS and GOODS-N, respectively, within $0\farcs7$. We find two matches with the COSMOS2015 catalog, at $\sim0\farcs4$. However, based on a visual inspection, there is no hint of emission at either 3 or 10\,GHz for these sources, indicating they are likely to be spurious. We find 37 matches within $0\farcs7$ between $\geq3\sigma$ peaks in the 34\,GHz GOODS-N radio map without radio counterparts and the 3D-HST catalog. In addition to a visual inspection, we stack these sources at 1.4\,GHz, but find no detection in the stack, and place a $3\sigma$ upper limit on the average 1.4\,GHz emission of $S_{1.4} \leq 2.7\,\mu\text{Jy\,beam}^{-1}$, which corresponds to a highly inverted $\alpha^{1.4}_{34} \gtrsim 0.6$. This further substantiates that the majority of the low-S/N peaks identified at 34\,GHz are likely to be spurious.

We summarize the radio properties of the eighteen sources detected at 34\,GHz in Table \ref{tab:coldz_sample}, and show postage stamps on top of \emph{Hubble Space Telescope} images in Figure \ref{fig:sources}. The tabulated flux densities are further corrected for flux boosting, as outlined in Appendix \ref{app:fluxboosting}. For a source detected at $3\sigma$ ($4\sigma$), the typical correction factor is $20\%$ ($10\%$), whereas at $\text{SNR}\gtrsim5$, the effects of flux boosting are found to be negligible.

\renewcommand{\arraystretch}{1.2}
\begin{deluxetable*}{lcccccccc}

\tabletypesize{\footnotesize}
\tablecolumns{9}
\tablewidth{\textwidth}
	
\tablecaption{Radio Properties of the 34\,GHz-selected COLD$z$ sample}

\tablehead{
	\colhead{\textbf{ID}} &
	\colhead{\textbf{RA}} &
	\colhead{\textbf{Dec}} &
	\colhead{\textbf{$z$\tablenotemark{a}}} &
	\colhead{\textbf{$S_{1.4}$}\tablenotemark{b}} &
	\colhead{\textbf{$S_{3}$}} &
	\colhead{\textbf{$S_{5}$}} &
	\colhead{\textbf{$S_{10}$}} &
	\colhead{\textbf{$S_{34}$}\tablenotemark{c}}
    }

\startdata

& [deg] & [deg] & & [$\mu$Jy] & [$\mu$Jy] & [$\mu$Jy] & [$\mu$Jy] & [$\mu$Jy] \\ \tableline \vspace{-1.0ex} \\

COLDz-cont-COS-1 & 150.093436 & 2.600324 & 0.89 & $<60.0$ & $8.8\pm0.6$ & - & $5.4\pm0.5$ & $3.2\pm1.4$ \\
COLDz-cont-COS-2\tablenotemark{d} & 150.086281 & 2.588947 & 5.30 & $<60.0$ & $21.9\pm0.6$ & - & $9.1\pm0.4$ & $6.8\pm1.3$ \\
COLDz-cont-COS-3 & 150.086022 & 2.600442 & $0.98 \pm 0.01$ & $272.0\pm13.0$ & $159.1\pm0.6$ & - & $78.2\pm0.5$ & $23.6\pm1.4$ \\
COLDz-cont-COS-4\tablenotemark{e} & 150.075874 & 2.582339 & 2.48 & $<60.0$ & $17.8\pm0.6$ & - & $7.6\pm0.5$ & $4.3\pm1.4$ \\
COLDz-cont-COS-5 & 150.097512 & 2.602361 & 1.00 & $<60.0$ & $27.4\pm0.6$ & - & $11.3\pm0.5$ & $6.3\pm1.5$ \\
COLDz-cont-GN-1 & 189.318274 & 62.253407 & 0.56 & $175.4\pm5.9$ & - & $153.0\pm3.7$ & $133.0\pm12.9$ & $72.3\pm4.3$ \\
COLDz-cont-GN-2\tablenotemark{f} & 189.251020 & 62.152715 & $1.61_{-0.01}^{+0.27}$ & $297.4\pm10.1$ & - & $114.2\pm5.3$ & - & $25.4\pm6.4$ \\
COLDz-cont-GN-3 & 189.247233 & 62.309206 & $2.00_{-0.01}^{+0.02}$ & $4600.8\pm78.0$ & - & $1106.0\pm6.0$ & - & $42.7\pm7.6$ \\
COLDz-cont-GN-4 & 189.222491 & 62.194332 & 1.27 & $74.4\pm7.6$ & - & $19.7\pm3.5$ & $17.4\pm2.1$ & $15.3\pm5.7$ \\
COLDz-cont-GN-5 & 189.220371 & 62.245561 & 0.32 & $214.0\pm7.8$ & - & $188.1\pm3.5$ & $191.5\pm3.5$ & $185.7\pm11.4$ \\
COLDz-cont-GN-6 & 189.193081 & 62.234673 & 0.96 & $278.2\pm9.4$ & - & $177.7\pm3.5$ & $138.9\pm2.9$ & $79.4\pm5.8$ \\
COLDz-cont-GN-7\tablenotemark{g} & 189.191979 & 62.246903 & $2.95_{-0.02}^{+0.06}$ & $103.9\pm3.7$ & - & $42.6\pm8.3$ & $35.1\pm2.9$ & $15.4\pm5.8$ \\
COLDz-cont-GN-8 & 189.184957 & 62.192532 & 1.01 & $1792.0\pm76.2$ & - & $963.0\pm6.3$ & $449.8\pm5.8$ & $317.8\pm8.2$ \\
COLDz-cont-GN-9 & 189.176004 & 62.262621 & 0.86 & $185.0\pm7.2$ & - & $46.3\pm4.3$ & $39.0\pm2.8$ & $15.6\pm5.1$ \\
COLDz-cont-GN-10 & 189.175494 & 62.225399 & 2.02 & $461.3\pm6.9$ & - & $137.8\pm3.5$ & $67.0\pm2.2$ & $30.6\pm5.7$ \\
COLDz-cont-GN-11\tablenotemark{h} & 189.143948 & 62.211490 & 1.22 & $188.5\pm7.7$ & - & $59.8\pm3.4$ & $31.5\pm2.4$ & $17.6\pm6.5$ \\
COLDz-cont-GN-12 & 189.297011 & 62.225238 & 2.00 & $126.0\pm6.3$ & - & $46.6\pm8.1$ & $16.3\pm2.6$ & $18.1\pm5.4$ \\
COLDz-cont-GN-13 & 189.193158 & 62.274894 & 0.50 & $403.0\pm14.2$ & - & $95.9\pm12.4$ & $26.0\pm4.8$ & $18.0\pm5.1$

\enddata

\tablenotetext{a}{Uncertainties are quoted on the photometric redshifts only.}
\tablenotetext{b}{Where applicable, $5\sigma$ upper limits are quoted for the COSMOS field, based on the typical RMS in the 1.4\,GHz map \citep{schinnerer2007,schinnerer2010}.}
\tablenotetext{c}{Flux densities are corrected for flux boosting (Appendix \ref{app:fluxboosting}). The flux densities listed for COLD$z$-cont-COS-2 and COLD$z$-cont-COS-4 are predominantly due to the combination of a bright CO-line and dust continuum emission (Section \ref{sec:linesubstraction}).}
\tablenotetext{d}{Detected in CO(2-1) emission in the COLD$z$ survey as COLDz.COS.0 \citep{pavesi2018}; also known as AzTEC-3.}
\tablenotetext{e}{Detected in CO(1-0) emission in the COLD$z$ survey as COLDz.COS.2 \citep{pavesi2018}.}
\tablenotetext{f}{Identified as GN-16 in the submillimeter observations by \citet{pope2005}}
\tablenotetext{g}{Identified as GN-12 by \citet{pope2005}.}
\tablenotetext{h}{Identified as GN-26 by \citet{pope2005}; has a CO(2-1)-based redshift from \citet{frayer2008}.}

\label{tab:coldz_sample}
\end{deluxetable*}
\renewcommand{\arraystretch}{1.0}

\section{Multi-Wavelength Properties}
\label{sec:multiwavelength}

\subsection{Multi-wavelength Counterparts and Redshifts}

Across the combined COSMOS and GOODS-N 34\,GHz mosaics, we identify a total of 18 robust high-frequency continuum detections. In this section, we detail the association of multi-wavelength counterparts to this radio-selected sample. For the five COSMOS sources, we follow the cross-matching procedure from \citet{algera2020b}. We first match the 34\,GHz continuum detections to the Super-deblended catalog containing far-infrared photometry, exploring matching radii up to $0\farcs9$. However, we find all five sources can be matched to Super-deblended counterparts within $0\farcs3$, where we expect a negligible number of false associations. We further find that the five sources can be additionally cross-matched to galaxies in the COSMOS2015 catalog, similarly within $0\farcs3$. We subsequently cross-match with the A2COSMOS and A3COSMOS ALMA catalogs, finding a single match at $0\farcs1$ that appears in both catalogs, and adopt the photometry from the former. We additionally match with the robust and tentative catalog of COLD$z$ CO-emitters from \citet{pavesi2018} and recover two matches within $0\farcs3$ within the robust set of blind CO-detections.\footnote{As such, the observed 34\,GHz continuum emission may be in part due to these bright CO-lines. We investigate this in Section \ref{sec:linesubstraction}.}

We then extract the optimal photometric or spectroscopic redshift from these catalogs. We prioritize redshifts in the following order: 1) a spectroscopic value based on a COLD$z$ CO-line, 2) a spectroscopic redshift within the Super-deblended catalog and 3) a photometric redshift within the COSMOS2015 catalog. In total, we find that 4 out of 5 COLD$z$ COSMOS detections have a spectroscopically confirmed redshift, while the remaining source has a well-constrained photometric redshift measurement (COLD$z$-cont-COS-3 at $z=0.98 \pm 0.01$). We note that one of our 34\,GHz continuum detections (COLD$z$-cont-COS-2) is the well-studied submillimeter galaxy AzTEC.3 at $z=5.3$, which is additionally detected in CO-emission in the COLD$z$ survey. For AzTEC.3, we further compile additional available ALMA continuum photometry at $230$ and $300\,$ GHz from \citet{pavesi2016}. \\

For the GOODS-N field, we follow a similar procedure. We first cross-match the COLD$z$ continuum detections with the 3D-HST survey \citep{brammer2012,skelton2014,momcheva2016}, adopting the radio positions at 1.4\,GHz from \citet{owen2018}, as these are of higher signal-to-noise than the 34\,GHz data, and as such are less susceptible to the local noise properties. We find that all 13 radio sources have counterparts in the 3D-HST survey within $0\farcs3$, where no spurious matches are expected. We subsequently cross-match with the Super-deblended catalog, and find that all but one of the 34\,GHz continuum sources have a counterpart at far-infrared wavelengths. All 12 cross-matches have a Super-deblended counterpart within $0\farcs3$, while there are no further matches within $3\farcs0$ of the single unmatched entry. We further find 3 cross-matches within $0\farcs1$ with the catalog of submillimeter-selected galaxies from \citet{pope2005}, which we identify with GN12, GN16 and GN26, the latter of which has a spectroscopically measured redshift of $z=1.22$ based on a CO(2-1) detection from \citet{frayer2008}. Additional matching with the COLD$z$ catalog of line emitters in the GOODS-N field does not result in further matches. As such, the $z=5.3$ submillimeter galaxy GN10 \citep{daddi2009,riechers2020}, detected in COLD$z$ as the brightest CO-emitter \citep{pavesi2018}, remains undetected in deep 34\,GHz imaging. Upon performing photometry at the known position of GN10, we determine a peak brightness of $S_{34} = 11.0 \pm 5.6\,\mu\text{Jy\,beam}^{-1}$ ($\approx2\sigma$), placing it well below our survey detection limit.

We compile the optimal redshifts for the 13 GOODS-N 34\,GHz continuum sources in a similar manner as for the COSMOS field. In order of priority, we adopt a spectroscopic redshift from the Super-deblended catalog, or the best redshift from 3D-HST \citep{momcheva2016}, the latter being either a spectroscopic redshift from \emph{HST}/GRISM, or a photometric redshift from \citet{skelton2014}. We further overwrite a single spectroscopic value for COLD$z$-cont-GN-10 at $z=4.424$ with the updated value of $z=2.018$, based on the discussion in \citet{murphy2017}. Overall, 10/13 sources have a spectroscopically confirmed redshift, while the remaining 3 sources have a well-constrained photometric redshift.

\subsection{Spectral Energy Distributions}
We use spectral energy distribution fitting code {\sc{magphys}} \citep{dacunha2008,dacunha2015} to determine the physical properties of the  34\,GHz continuum detections. {\sc{magphys}} adopts an energy balance technique to couple the stellar emission at ultra-violet to near-infrared wavelengths to thermal dust emission at longer wavelengths, and as such models the ultra-violet to far-infrared SED in a self-consistent way. This is additionally useful for sources lacking far-infrared photometry, as in this case shorter wavelengths will still provide constraints on the total dust emission. While {\sc{magphys}} is also capable of modelling the radio spectrum of star-forming galaxies, we do not utilize any observations at wavelengths beyond $1.3\,$mm in our SED-fitting procedure, as any emission from an active galactic nucleus at radio wavelengths is not incorporated in the fitting. We additionally add a 10\% uncertainty in quadrature to the catalogued flux density uncertainties bluewards of \emph{Spitzer}/MIPS $24\mu$m, following e.g., \citet{battisti2019}. This accounts for uncertainties in the photometric zeropoints, and further serves to guide the fitting into better constraining the far-infrared part of the spectral energy distribution. Via {\sc{magphys}}, we obtain several physical properties for our galaxies, including far-infrared luminosities, star-formation rates and stellar masses. We show the fitted spectral energy distributions for all eighteen COLD$z$ continuum detections in Figure \ref{fig:seds} in Appendix \ref{app:seds}, and tabulate their physical parameters in Table \ref{tab:magphys}.

\subsection{X-ray and Mid-infrared AGN Signatures}

Given that the COLD$z$ continuum detections in the COSMOS field are all identified in the COSMOS-XS survey, we adopt the results from \citet{algera2020b}, who match the COSMOS-XS 3\,GHz continuum sources with the X-ray catalog from \citet{marchesi2016}. None of the five COLD$z$ COSMOS continuum sources, however, have counterparts in X-ray emission within a separation of $1\farcs4$, and as such we calculate upper limits on their X-ray luminosity adopting $\Gamma = 1.8$. However, as the resulting limits are of modest depth, we cannot state definitively whether the COLD$z$ COSMOS sources are X-ray AGN. We note, however, that all of them fall below the typical X-ray emission seen in submillimeter galaxies \citep{alexander2005}, which in turn are thought to predominantly be star-formation dominated. We additionally adopt the criteria from \citet{donley2012} in order to identify AGN at $z\leq2.7$ through the mid-infrared signature ascribed to a dusty torus surrounding the accreting black hole, and similarly find no signature of AGN-related emission at mid-infrared wavelengths. We limit ourselves to this redshift range, as at higher redshifts the \citet{donley2012} criteria are susceptible to false positives in the form of dusty star-forming galaxies (e.g., \citealt{stach2019}).

For the GOODS-N field, we find 11/13 matches with the X-ray catalog from \citet{xue2016}, within a separation of $1\farcs0$. At this matching radius, no false identifications are expected. We compare the X-ray luminosities with the emission expected from star-formation, adopting the relations from \citet{symeonidis2014}. This comparison classifies nine sources as AGN based on their X-ray emission, despite two of these sources having [0.5-8]\,keV X-ray luminosities below the typical threshold for AGN of $L_X = 10^{42}\,\text{erg\,s}^{-1}$. We additionally find that only a single X-ray detected source exhibits mid-infrared colors that place it within the \citet{donley2012} wedge. Overall, we conclude that the majority of 34\,GHz continuum detections in the GOODS-N field are X-ray AGN. Due to the COSMOS X-ray data being comparatively shallower, we cannot assess definitely whether the five 34\,GHz continuum sources in the COSMOS field are similarly AGN based on their X-ray emission. However, a lower AGN fraction in COSMOS is expected, as its radio observations are significantly deeper than in GOODS-N, and the incidence rate of AGN is a strong function of radio flux density \citep{smolcic2017b,algera2020b}. Regardless, if X-ray AGN are present in the COLD$z$/COSMOS detections, they are unlikely to be dominating the spectral energy distribution.


\begin{figure*}
    \centering
    \includegraphics[width=\textwidth]{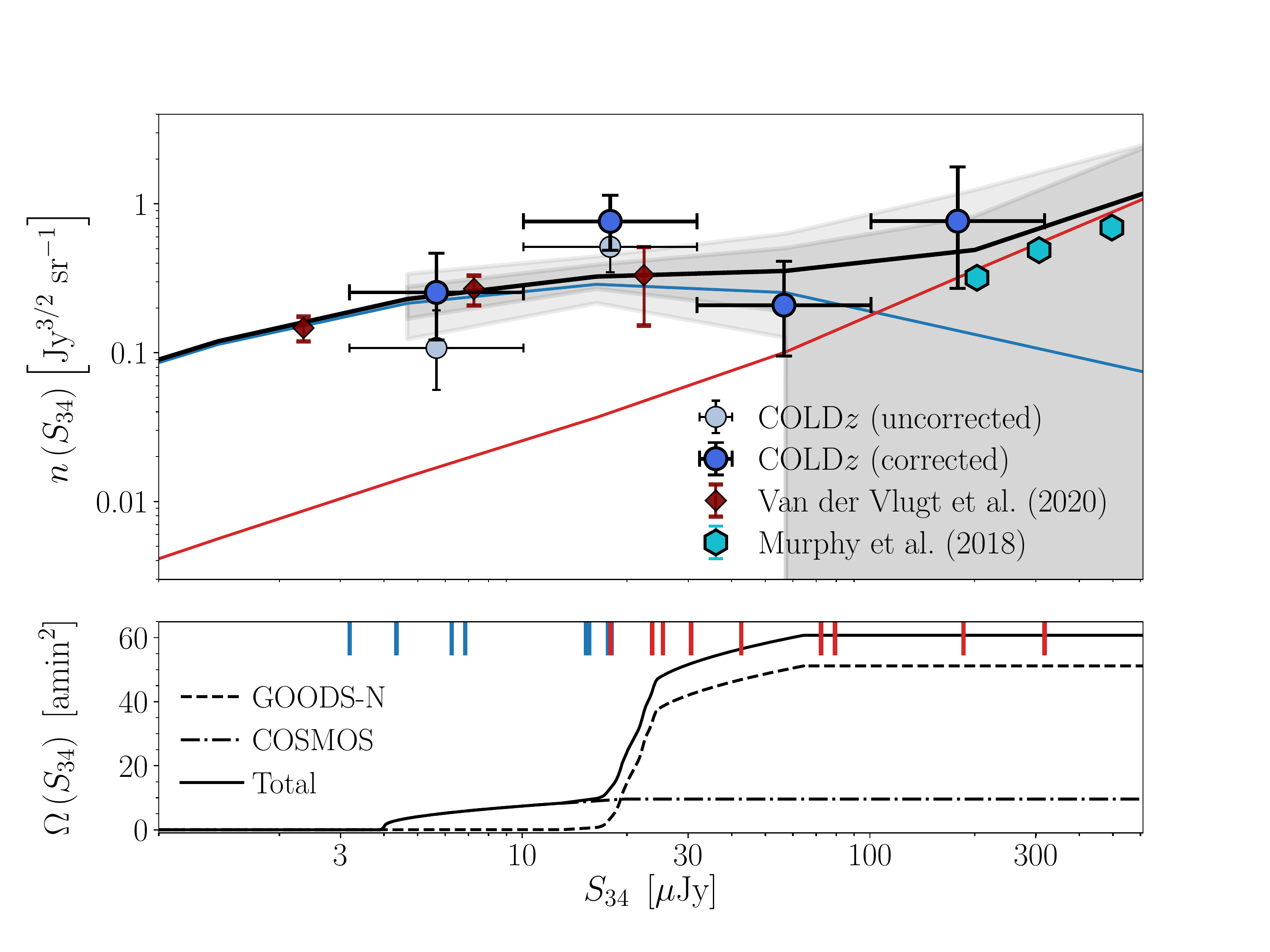}
    \caption{\textbf{Upper:} Completeness-corrected Euclidean-normalized radio source counts at 34\,GHz across the combined COSMOS and GOODS-N mosaics. The purple diamonds show the 10\,GHz counts from \citet{vandervlugt2020}, and the cyan hexagons represent the 28.5\,GHz counts from \citet{murphy2018}, both rescaled to 34\,GHz assuming $\alpha=-0.70$. The blue and red lines indicate the scaled source counts for star-forming galaxies and AGN from the \citet{bonaldi2019} simulations at 20\,GHz, respectively. The dark (light) grey regions indicate the expected $1\sigma$ ($2\sigma$) level of cosmic variance, which drops to zero at low flux densities when our effective area is zero. \textbf{Lower:} The effective area of the COLD$z$ observations as a function of flux density. For $S_{34} \lesssim 20\,\mu$Jy, we are restricted to the relatively small field-of-view of the COLD$z$ COSMOS observations. The blue (red) vertical bars indicate the (deboosted) 34\,GHz flux density at which a star-forming galaxy (radio AGN) is detected (Section \ref{sec:radioproperties}). Despite the relatively large uncertainties, our 34\,GHz number counts are in good agreement with (rescaled) measurements and predictions in the literature.}
    \label{fig:numbercounts}
\end{figure*}

\begin{deluxetable}{ccccc}

\tabletypesize{\footnotesize}
\tablecolumns{5}
\tablewidth{0.4\textwidth}
	
\tablecaption{Euclidean Number Counts at 34\,GHz}

\tablehead{
	\colhead{{$S_\text{centre}$}} &
	\colhead{{$S_\text{low}$}} &
	\colhead{{$S_\text{high}$}} &
	\colhead{{$n_\text{uncorr}(S)$}} &
	\colhead{{$n_\text{corr}(S)$}}
    }

\startdata

[$\mu$Jy] & [$\mu$Jy] & [$\mu$Jy] & [$\text{Jy}^{3/2}\,\text{sr}^{-1}$] & [$\text{Jy}^{3/2}\,\text{sr}^{-1}$] \\ 

\tableline \vspace{-1.0ex} \\

5.7 & 3.2 & 10.1 & $0.108_{-0.051}^{+0.085}$ & $0.254_{-0.132}^{+0.212}$ \\
17.9 & 10.1 & 31.8 & $0.514_{-0.168}^{+0.235}$ & $0.763_{-0.274}^{+0.382}$ \\
56.6 & 31.8 & 100.6 & $0.209_{-0.114}^{+0.203}$ & $0.209_{-0.114}^{+0.203}$ \\
178.8 & 100.6 & 317.8 & $0.765_{-0.494}^{+1.010}$ & $0.765_{-0.494}^{+1.010}$ 

\enddata

\tablenotetext{}{(1) Central flux density of the bin, (2), (3) Lower and upper flux density of the bin, (4) Number counts uncorrected for incompleteness, (5) Number counts corrected for incompleteness.}

\label{tab:counts}
\end{deluxetable}

\section{34 GHz Source Counts}
\label{sec:sourcecounts}

Radio number counts, while historically used as a probe for the cosmology of the Universe, remain a useful tool for comparing surveys, in addition to visualizing the onset of different radio populations. At the bright end ($S_{1.4} \gg 1\,$mJy), the Euclidean number counts decline smoothly towards lower flux densities, and are dominated by luminous radio AGN (e.g., \citealt{condon1984}). At 1.4\,GHz, the number counts show a flattening at $S_{1.4} \approx 1\,$mJy, believed to be the advent of star-forming galaxies and radio-quiet AGN as the dominant radio populations (e.g., \citealt{rowan-robinson1993, seymour2004,padovani2009,smolcic2017b}). For a typical spectral index of $\alpha = -0.70$, this flattening should arise around $S_{34}\approx100\,\mu$Jy, and hence is covered in the range of flux densities probed in this work. We note that these flux densities may be ``contaminated'' by thermal emission from dust, or in the case of two COSMOS sources, by bright CO-emission, which are typically not an issue in low-frequency radio source counts. However, while we correct for this when examining the radio spectra of the star-forming COLD$z$ continuum detections in detail (Section \ref{sec:decomposition}), here we compute the number counts based on the raw observed flux densities.

As we adopt deep radio observations as prior positions for possible 34-GHz continuum sources, we expect our sample to be fully reliable, i.e., not to contain any spurious detections. However, our sample may still be incomplete, in particular as a result of the RMS of our radio maps increasing rapidly towards the image edges due to the enhanced primary beam attenuation (c.f., Figure \ref{fig:rmsmaps}). In turn, in these regions we may miss faint 34\,GHz continuum sources that would have been observed had the RMS been constant across our field-of-view. As such, we adopt the fractional incompleteness as a function of radio flux density as determined from inserting mock sources into our radio maps (Appendix \ref{app:completeness}). Denoting the completeness at flux density $S_\nu$ for field $i$ as $f_i(S_\nu)$, the correction factor per field is simply $C_i = f_i^{-1}(S_\nu)$. For multiple fields, we then adopt the full completeness to be 

\begin{align}
    C(S_\nu) = \frac{\sum_i f_i^{-1}(S_\nu) \Omega_i(S_\nu)}{\sum_i \Omega_i(S_\nu)}\ ,
    \label{eq:completeness}
\end{align}

\noindent where $\Omega_i(S_\nu)$ denotes the area in arcmin$^2$ across which a source of flux density $S_\nu$ can be detected in field $i$ at $\geq 3\sigma$ significance. As such, the overall completeness is the area-weighted average of the completeness in the COSMOS and GOODS-N fields.

We present the completeness-corrected Euclidean-normalized number counts at 34\,GHz in Figure \ref{fig:numbercounts}, and tabulate the results in Table \ref{tab:counts}. The number counts combine the continuum detections across both the COSMOS and GOODS-N fields, using the same area-weighting that is adopted for the completeness calculation (Equation \ref{eq:completeness}). The uncertainties on the individual points constitute the combination of the error on the counting statistics from \citet{gehrels1986} -- which are more appropriate than simple Poissonian errors for bins with few sources -- and the error on the completeness. Cosmic variance is not included in the uncertainties, but its magnitude is discussed below.

While the high-frequency ($\nu \gtrsim 30\,$GHz) radio sky has been explored at the milliJansky level (e.g., \citealt{mason2009}), the COLD$z$ survey provides the first constraints on the 34\,GHz number counts in the regime where star-forming galaxies are expected to emerge as the dominant population, complicating any direct comparisons to the literature. At the bright end of our 34\,GHz observations, we compare with \citet{murphy2018}, who perform a stacking analysis in \emph{Planck} observations at 28.5\,GHz, based on priors at 1.4\,GHz from NVSS \citep{condon1998}. We rescale the number counts from 28.5\,GHz to 34\,GHz adopting $\alpha=-0.70$, and find them to be in agreement with the COLD$z$ observations at $S_{34}\gtrsim200\,\mu$Jy. 

As no 34\,GHz observations with a similar sensitivity to COLD$z$ exist in the literature, the next best solution is to compare to the highest-frequency number counts available, and scale the counts to 34\,GHz. For this, we adopt the recent 10\,GHz number counts from \citet{vandervlugt2020}, as part of the COSMOS-XS survey, which we scale to 34\,GHz via $\alpha=-0.70$. As the parent survey constitutes a very deep single pointing ($\sim0.40\,\mu$Jy across $\sim30\,\text{arcmin}^2$), these data probe down to slightly fainter flux densities than those probed in this work. However, within the flux density range in common, we find that the number counts are in good agreement, despite the inherent uncertainties associated with the required frequency scaling.

While observationally little is known about the faint high-frequency radio sky, this is no less true for simulations. Modelling of the radio sky has predominantly been performed at low frequencies, with simulations by \citet{wilman2008} and more recently by \citet{bonaldi2019} only extending to 18 and 20\,GHz, respectively. As such, we again invoke a frequency scaling to 34\,GHz, adopting as before a spectral index of $\alpha=-0.70$. We focus on the recent 20\,GHz simulations by \citet{bonaldi2019}, who model the radio population for two distinct classes of sources: star-forming galaxies that follow the far-infrared/radio correlation, and radio AGN that show a strong excess in radio power compared to this correlation. We show both the individual and combined contributions of the two populations in Figure \ref{fig:numbercounts}, and find that the simulations predict that below $S_{34} \lesssim 100\,\mu$Jy star-forming galaxies should make up the bulk of the radio population. Our measurements are in good agreement with the simulated number counts, indicating that a scaling from 20 to 34\,GHz with a fixed spectral index is likely to be appropriate. We caution, however, that the uncertainties on our number counts are large as a result of the small number of sources detected in the 34\,GHz continuum maps.

Furthermore, cosmic variance constitutes an additional uncertainty on our number counts, as our radio observations probe a relatively small field-of-view. We show the area probed as a function of flux density in the bottom panel of Figure \ref{fig:numbercounts}. Out to $S_{34} \lesssim 15\,\mu$Jy, the total area is dominated by the deeper COSMOS mosaic, and accounts for approximately $10\,\text{arcmin}^2$. At flux densities above $S_{34}\gtrsim 30\,\mu$Jy, our field-of-view is increased to approximately $60\,\text{arcmin}^2$, as now sources can be detected across the full GOODS-N mosaic as well. We quantify the magnitude of cosmic variance via the \citet{bonaldi2019} simulations, following \citet{algera2020b}. Briefly, for each of the flux density bins adopted in our computation of the 34\,GHz number counts, we determine the effective area of the COLD$z$ survey in which sources in the given bin can be detected. We then sample 200 independent circular regions of equivalent area from the 20\,GHz Bonaldi simulations, accounting for the flux density scaling with $\alpha=-0.70$. We determine the number counts for all independent areas, and compute the 16-84$^\text{th}$ and 5-95$^\text{th}$ percentiles, which we adopt to be the $1$ and $2\sigma$ uncertainties due to cosmic variance, shown as the dark and light grey regions in Figure \ref{fig:numbercounts}, respectively. We note that such a calculation of the cosmic variance encapsulates two effects: at low flux densities we have a relatively small field-of-view, which naturally increases the magnitude of cosmic variance. At large flux densities, our field of view constitutes the full $\sim60\,\text{arcmin}^2$, but due to the relative paucity of bright radio sources, cosmic variance similarly constitutes an appreciable uncertainty. The typical magnitude of cosmic variance between $5 \lesssim S_{34} \lesssim 60\,\mu$Jy induces an additional uncertainty on the number counts of $\sim0.1-0.2\,$dex -- comparable to the Poissonian uncertainties -- although this value rapidly increases for higher flux densities. Overall, we therefore conclude that our number counts are in good agreement with both observed and simulated counts from the literature at lower-frequencies.


\section{34\,GHz Continuum Source Properties}
\label{sec:radioproperties}

\begin{figure*}[!t]
    \centering
    \includegraphics[width=\textwidth]{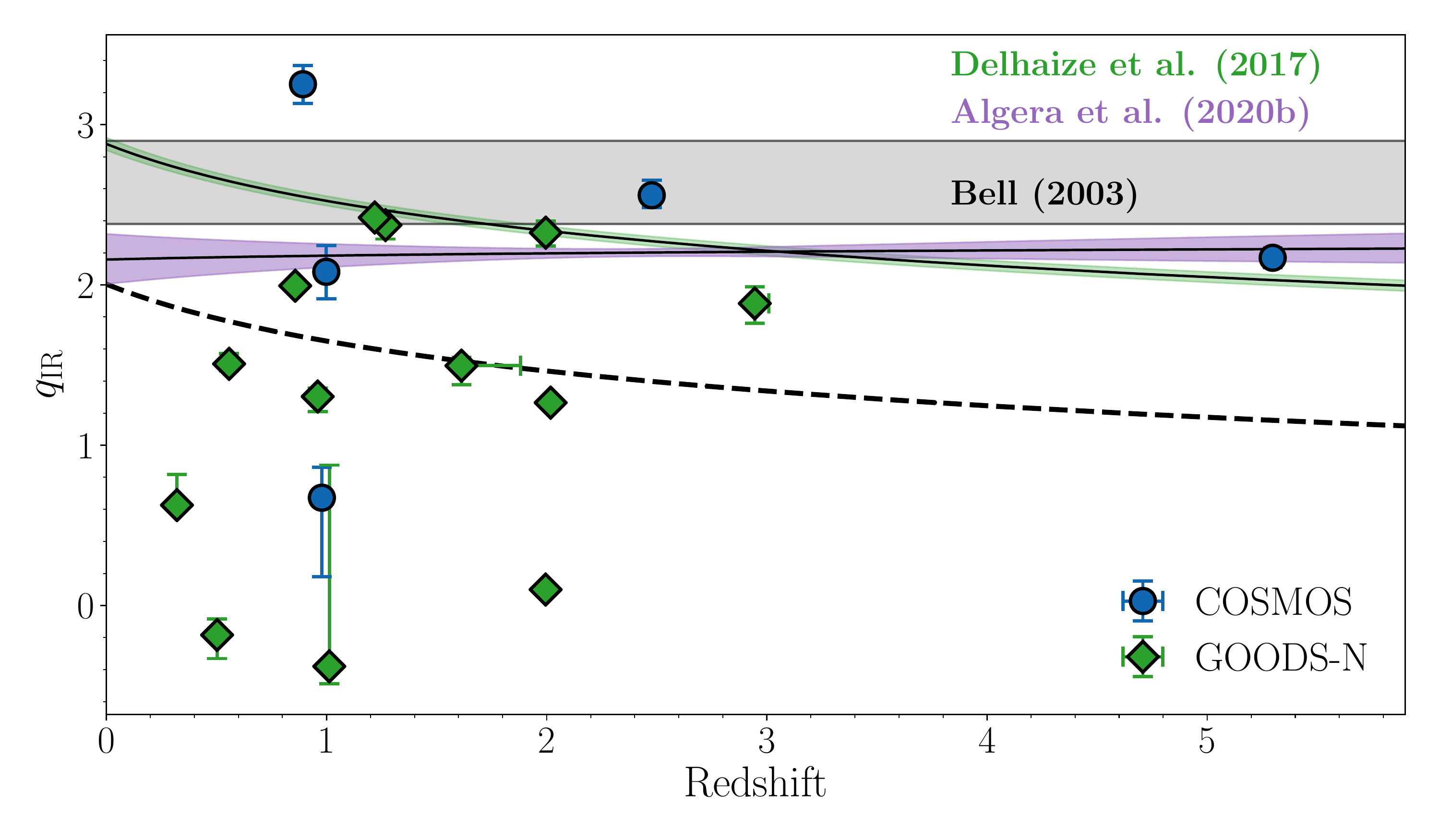}
    \caption{The far-infrared/radio correlation for the COLD$z$ 34\,GHz continuum detections. Three literature results for the correlation are shown: the local value from \citet{bell2003}, the correlation for 3\,GHz-selected galaxies from \citet{delhaize2017} and the correlation for submillimeter galaxies from \citet{algera2020a}. The threshold for identifying radio-excess AGN is shown via the dashed curve, which is the FIRRC from \citet{delhaize2017} minus $2.5\times$ the intrinsic scatter. We identify half of the 34\,GHz continuum sample (9/18 sources) as radio AGN.}
    \label{fig:FIRRC}
\end{figure*}

\subsection{Radio AGN}
\label{sec:radioagn}
In the local Universe, the existence of a linear correlation between the total far-infrared and radio emission of star-forming galaxies has been well-established \citep{yun2001,bell2003}. This far-infrared/radio correlation has been shown to hold over a wide range of luminosities, from dwarf galaxies to dust-obscured starbursts \citep{bell2003}. The correlation is commonly expressed via parameter $q_\text{IR}$, first introduced by \citet{helou1985}, and defined as

\begin{align}
	q_\text{IR} = \log_{10} \left( \frac{L_\text{IR}}{3.75 \times 10^{12}\text{ W}} \right) - \log_{10} \left( \frac{L_\text{1.4}}{\text{W Hz}^{-1}} \right)  \ .
	\label{eq:qIR}
\end{align}

\noindent Here $L_\text{IR}$ represents the $8-1000\,\mu$m luminosity, and $L_{1.4}$ is the $K$-corrected radio-luminosity at rest-frame 1.4\,GHz, computed via the closest observed-frame flux density and the measured radio spectral index. A number of recent, radio-selected studies of the far-infrared/radio correlation have found it to evolve with cosmic time (e.g., \citealt{delhaize2017,calistrorivera2017,read2018,ocran2020}). While the origin of this evolution is debated, and may be the result of residual AGN contamination \citep{molnar2018}, or due to the different physical conditions in massive star-forming galaxies at high-redshift compared to local sources \citep{algera2020a,delvecchio2020}, we adopt an evolving far-infrared/radio correlation to identify radio-AGN among the eighteen 34\,GHz detections. In particular, we adopt the correlation determined for a 3\,GHz selected sample by \citet{delhaize2017}, and follow \citet{algera2020b} by identifying galaxies as radio AGN if they fall below the correlation at the $2.5\sigma$ level.

We show the far-infrared/radio correlation for the eighteen 34\,GHz continuum detections as a function of redshift in Figure \ref{fig:FIRRC}. In total, half of the COLD$z$ sample are identified as radio AGN, comprising eight AGN in GOODS-N, and one in COSMOS (Table \ref{tab:magphys}). The fact that radio AGN make up a large fraction of the bright radio population is evidenced by the relatively large contribution of AGN in the wider but shallower GOODS-N observations. Among the galaxies classified as star-forming, Figure \ref{fig:sources} suggests the existence of an extended tail at 34\,GHz in COS-2, at face value indicative of an AGN jet. However, such extended features are absent in the deeper 3\, and 10\,GHz observations of this source, and at these frequencies the galaxy is consistent with being a point source at $2''$ resolution. As the spectrum of jetted AGN tends to steepen towards higher frequencies (e.g., \citealt{mahatma2018}), and should therefore be detectable in the lower frequency ancillary radio data, we interpret the extended tail at 34\,GHz as simply being due to noise. As we adopt the peak brightness for COS-2, its flux density measurement is unlikely to be substantially affected by this noisy region in the radio map.

In total, 6/8 radio-AGN in the GOODS-N field are additionally classified as AGN through their strong X-ray emission. While constituting only a small number of sources, this is a relatively large fraction, as the overlap between radio AGN selected at lower frequencies and X-ray AGN is typically found to be small ($\lesssim30\%$, e.g., \citealt{smolcic2017b,delvecchio2017,algera2020b}). This difference, however, may in part be due to the deeper X-ray data available in GOODS-N, compared to the COSMOS field where these studies were undertaken.

\begin{figure*}
    \centering
    \hspace*{-0.2cm}
    \includegraphics[width=1.05\textwidth]{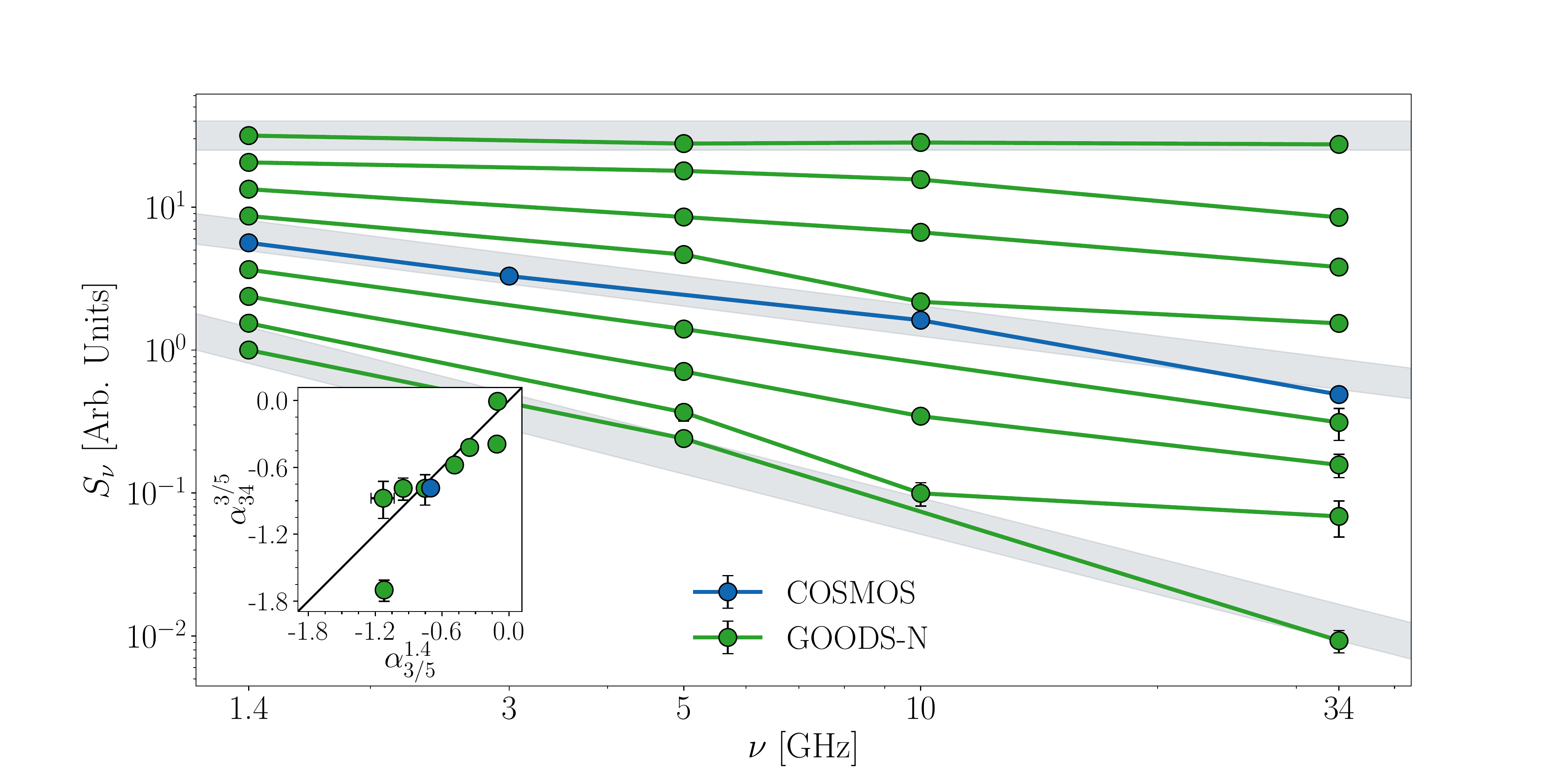} \hfill
    \caption{The radio spectra of the nine radio AGN identified in COSMOS (1) and GOODS-North (8). The sources have been arbitrarily normalized, and are arranged from the flattest to steepest $1.4-34\,$GHz spectral index. Note that the errorbars are typically smaller than the chosen plotting symbols. The grey bands indicate slopes of $\alpha=0.0, -0.70, -1.40$ (from top to bottom), comparable to the flattest, median and steepest $1.4-34\,$GHz spectral slopes we observe for the COLD$z$ AGN, and are shown for reference. The inset shows the $5-34\,$GHz spectral index versus the $1.4-5\,$GHz slope ($3-34\,$GHz and $1.4-3\,$GHz for COSMOS). The radio AGN show a variety of spectral slopes, with a median spectral index of $\alpha^{1.4}_{34} = -0.77 \pm 0.13$. However, the AGN exhibit relatively smooth radio spectra, with only two sources showing strong evidence for spectral curvature.}
    \label{fig:agn_spectra}
\end{figure*}

We further show the long-wavelength spectra of the nine radio AGN in Figure \ref{fig:agn_spectra}. The median $1.4-34\,$GHz spectral index of the sample equals $\alpha = -0.77 \pm 0.13$, which is consistent with the commonly assumed synchrotron slope of $\alpha = -0.70$. However, the modest sample still spans a wide range of spectral slopes between $1.4$ and $34\,$GHz, ranging from nearly flat ($\alpha = -0.04 \pm 0.02$) to steep ($\alpha = -1.47 \pm 0.06$). In addition, the AGN exhibit relatively smooth spectra, with the median spectral index between $1.4-3\ (\text{or}\ 5)\,$GHz of $\alpha^{1.4}_{3/5} = -0.70 \pm 0.12$ being consistent with the typical high-frequency slope of $\alpha^{3/5}_{34} = -0.79 \pm 0.15$. Only two AGN exhibit strong evidence for steepening of their radio spectra towards higher frequencies, although sources with strongly steepening spectra are more likely to be missed in a selection at high radio frequencies. Such spectral steepening is expected to occur due to synchrotron ageing losses increasing towards higher frequencies, and in turn relates to the age of the AGN (e.g., \citealt{carilli1991}).

\subsection{Radio Spectral Decomposition for Star-forming Galaxies}
\label{sec:decomposition}

Detecting radio free-free emission in high-redshift star-forming galaxies is challenging due to its expected faintness, and the presence of a radio AGN only further hinders the detection of this already elusive component in the radio spectrum. As such, we now turn our attention to the star-formation-powered sources detected in the COLD$z$ survey. The radio spectrum of star-forming galaxies is frequently assumed to be the superposition of two power laws arising from non-thermal synchrotron and thermal free-free emission \citep{condon1992,murphy2017,tabatabaei2017}. Denoting their spectral indices as $\alpha_\text{NT}$ and $\alpha_\text{FF}$, respectively, the radio flux density at a given frequency $\nu$ may be written as

\begin{align}
    S_\nu = S_{\nu_0}^\text{NT} \left( \frac{\nu}{\nu_0}\right)^{\alpha_\text{NT}} + S_{\nu_0}^\text{FF} \left( \frac{\nu}{\nu_0}\right)^{\alpha_\text{FF}} \ , \label{eq:spectrum1}
\end{align}

\noindent given a reference frequency $\nu_0$, as well as the thermal and non-thermal flux densities $S_{\nu_0}^\text{FF}$ and $S_{\nu_0}^\text{NT}$, respectively, evaluated at this frequency. We rewrite this equation by introducing the thermal fraction, defined through $f_{\nu_0}^\text{th} = S_{\nu_0}^\text{FF} / (S_{\nu_0}^\text{FF} + S_{\nu_0}^\text{NT})$, as is common in the literature (e.g., \citealt{condon1992,tabatabaei2017}). As such, we rewrite the radio spectrum as

\begin{align}
    S_\nu = \left( 1-f_{\nu_0}^\text{th} \right) S_{\nu_0} \left( \frac{\nu}{\nu_0}\right)^{\alpha_\text{NT}} + f_{\nu_0}^\text{th} S_{\nu_0} \left( \frac{\nu}{\nu_0}\right)^{-0.1} \ .
    \label{eq:spectrum2}
\end{align}

\noindent This further assumes the spectral index for thermal free-free emission is fixed at $\alpha_\text{FF} = -0.10$ \citep{condon1992,murphy2011}. We then determine the remaining free parameters, $f^\text{th}_{\nu_0}, \alpha_\text{NT}$ and $S_{\nu_0}$, using a Monte Carlo Markov Chain (MCMC) based fitting routine. In addition, we adopt an observer-frame frequency of 1.4\,GHz as the reference frequency $\nu_0$, which defines the frequency where the thermal fraction is normalized. Where necessary, we convert the thermal fraction from observed-frame frequency $\nu$ to a rest-frame frequency $\nu'$ via 

\begin{align}
    f_{\nu'}^\text{th} = \frac{ f^\text{th}_{\nu(1+z)} \left( \frac{\nu'}{\nu (1+z)}\right)^{-0.10}}{f^\text{th}_{\nu(1+z)} \left( \frac{\nu'}{\nu (1+z)}\right)^{-0.10} + (1 - f^\text{th}_{\nu(1+z)}) \left( \frac{\nu'}{\nu (1+z)}\right)^{\alpha_\text{NT}}} \ .
    \label{eq:scale_thermal}
\end{align}

\noindent We adopt flat priors in our fitting routine for the normalization $S_{\nu_0}$ and thermal fraction $f_\text{th}$, while we adopt Gaussian priors for the non-thermal spectral index (following, e.g., \citealt{linden2020}). For the former two parameters, we require that $-1\,\text{Jy} < S_{\nu_0} < 1\,\text{Jy}$ and $-0.5 < f_\text{th} < 1.5$, that is, we allow both the normalization and the thermal fraction to take on unphysical, negative values, as artificially bounding both to be greater than zero will not fully capture the uncertainties in the MCMC-sampling. In addition, allowing negative values in the thermal fraction will demonstrate the necessity for the thermal component, as well as limitations inherent to the simple model we adopt for the radio spectrum (see also \citealt{tabatabaei2017}).

The Gaussian priors adopted on the non-thermal spectral index are motivated by a degeneracy that manifests between the synchrotron slope and thermal fraction at low signal-to-noise. This degeneracy occurs because it is impossible to accurately distinguish between an overall flat spectrum as being due to dominant free-free emission, or an intrinsically shallow synchrotron slope. To partially alleviate this degeneracy, we adopt prior knowledge from the local Universe that the synchrotron spectral indices are typically distributed around $\alpha_\text{NT} \approx -0.85$ \citep{niklas1997,murphy2011}. In large radio-selected samples, the scatter around the typical low-frequency spectral index equals $0.3 - 0.5\,$dex \citep{smolcic2017a,calistrorivera2017,gim2019}, and is approximately Gaussian. While these spectral index measurements include both the synchrotron and free-free components, the low-frequency nature of these data ensure the radio fluxes are likely dominated by non-thermal emission, and as such, the overall variation in the spectral index constitutes a proxy for the scatter in the typical synchrotron spectral index. To encompass the full observed scatter in the synchrotron slopes, we therefore adopt a Gaussian prior on $\alpha_\text{NT}$ centered on a mean value of $-0.85$, with a scatter of $0.50$.

\subsection{Line and Dust Continuum Subtraction}
\label{sec:linesubstraction}

Prior to fitting the radio spectra of the star-forming COLD$z$ continuum detections, we need to ensure the 34\,GHz flux density is not contaminated by thermal emission from dust, or by strong line emission. In particular, the 34\,GHz flux density of the two COLD$z$ COSMOS sources detected in CO-emission (COS-2, a.k.a.\ AzTEC.3, and COS-4), may be boosted by their respective CO-lines. We re-extract the peak brightness of these two sources after removing the channels contaminated by the CO-emission, and, as a sanity check, repeat this for the three COSMOS sources that do not show any evidence for strong line emission. While for the latter the flux densities are unaffected by this procedure, we find line-uncontaminated flux densities for COS-2 and COS-4 of $S_{34} = 5.2 \pm 1.3\,\mu$Jy and $S_{34} = 3.0 \pm 1.4\,\mu$Jy, respectively, which are lower than the original catalogued flux densities by $\sim25\%$ and $\sim40\%$ (Table \ref{tab:coldz_sample}). In turn, this correction brings the 34\,GHz flux density from COS-4 below the formal detection limit ($\text{SNR}\approx2$). \\

The 34\,GHz continuum flux densities may further contain a contribution from thermal emission by dust, which, at least for local normal star-forming galaxies, is thought to dominate the radio spectrum beyond rest-frame $\nu \gtrsim 100-200\,$GHz \citep{condon1992}. To determine the extent of this contribution, we fit -- where available -- the far-infrared observations of our galaxy sample with both optically thin and thick modified blackbody spectra, and extrapolate the resulting dust SED to observed-frame 34\,GHz. We note that this methodology is rather sensitive to how well the global dust properties (e.g., temperature and emissivity) can be constrained, and as a result the predicted flux densities are quite uncertain. Nevertheless, we find that the 34\,GHz emission of COS-2 and COS-4 is likely to be dominated by emission from dust, with predicted dust contributions of $4.9_{-3.6}^{+6.1}\,\mu$Jy and $3.3_{-1.5}^{+7.2}\,\mu$Jy, respectively. As a result, the full 34\,GHz flux densities of these two sources are consistent with being powered by the combination of CO-emission and dust. As we probe rest-frame frequencies of $\nu' \approx210\,$GHz and $\nu' \approx120\,$GHz for COS-2 and COS-4, respectively, this finding is consistent with the typical model for the long-wavelength SED of star-forming galaxies \citep{condon1992}.

In what follows, we will discard COS-2 and COS-4 from our sample, as any remaining contribution from free-free or synchrotron emission to the measured 34\,GHz flux density is not statistically significant, such that no robust spectral decomposition for these two sources can be performed. One source in GOODS-N, GN-7 at $z=2.95$ ($\nu' \approx 130\,$GHz), may have $\sim25\%$ of its continuum flux density contaminated by dust emission. However, due to the aforementioned uncertainties in the fitting of the dust SED, we do not correct for this potential contribution. This analysis indicates that, even at low flux densities ($S_{34} \lesssim 25\,\mu$Jy), a 34-GHz-selected sample does not automatically yield a free-free-dominated population.

\subsection{The Radio Spectra of High-redshift Star-forming Galaxies}

\begin{figure*}[!t]
    \centering
    \includegraphics[width=0.95\textwidth]{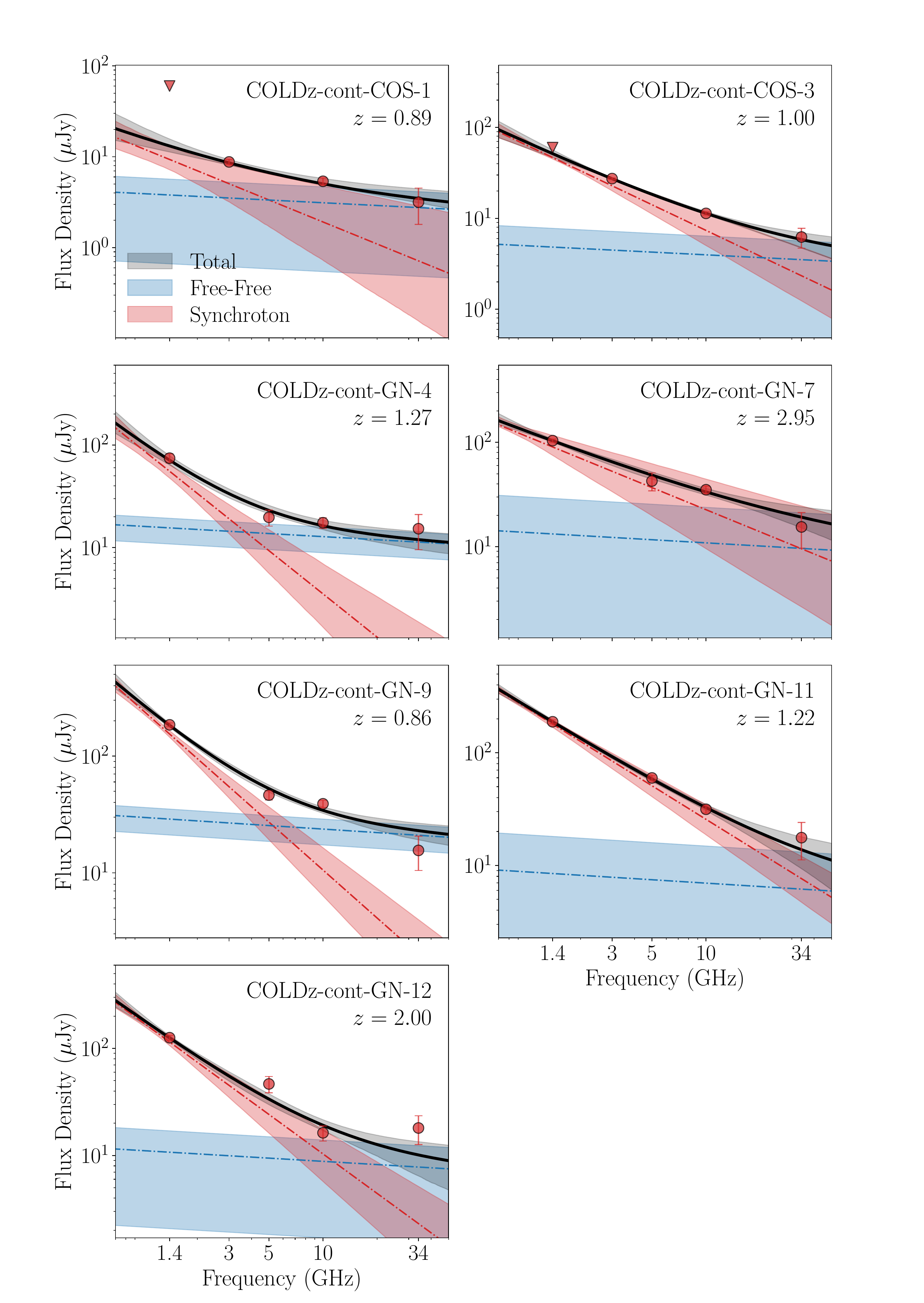}
    \caption{The radio spectra of the seven star-forming galaxies in the COSMOS (first 2 panels) and GOODS-North (last 5) fields. We show the decomposition of the spectra into their synchrotron and free-free components, with the shaded regions indicating the $1\sigma$ confidence region on the fits. We find the radio emission for four out of seven galaxies (COS-1, GN-4, GN-9 and GN-12) to be dominated by free-free emission at observed-frame 34\,GHz, whereas for the remainder only a relatively minor thermal contribution is preferred.}
    \label{fig:SFG_spectra}
\end{figure*}

\begin{figure*}[!t]
    \centering
    \includegraphics[width=1.00\textwidth]{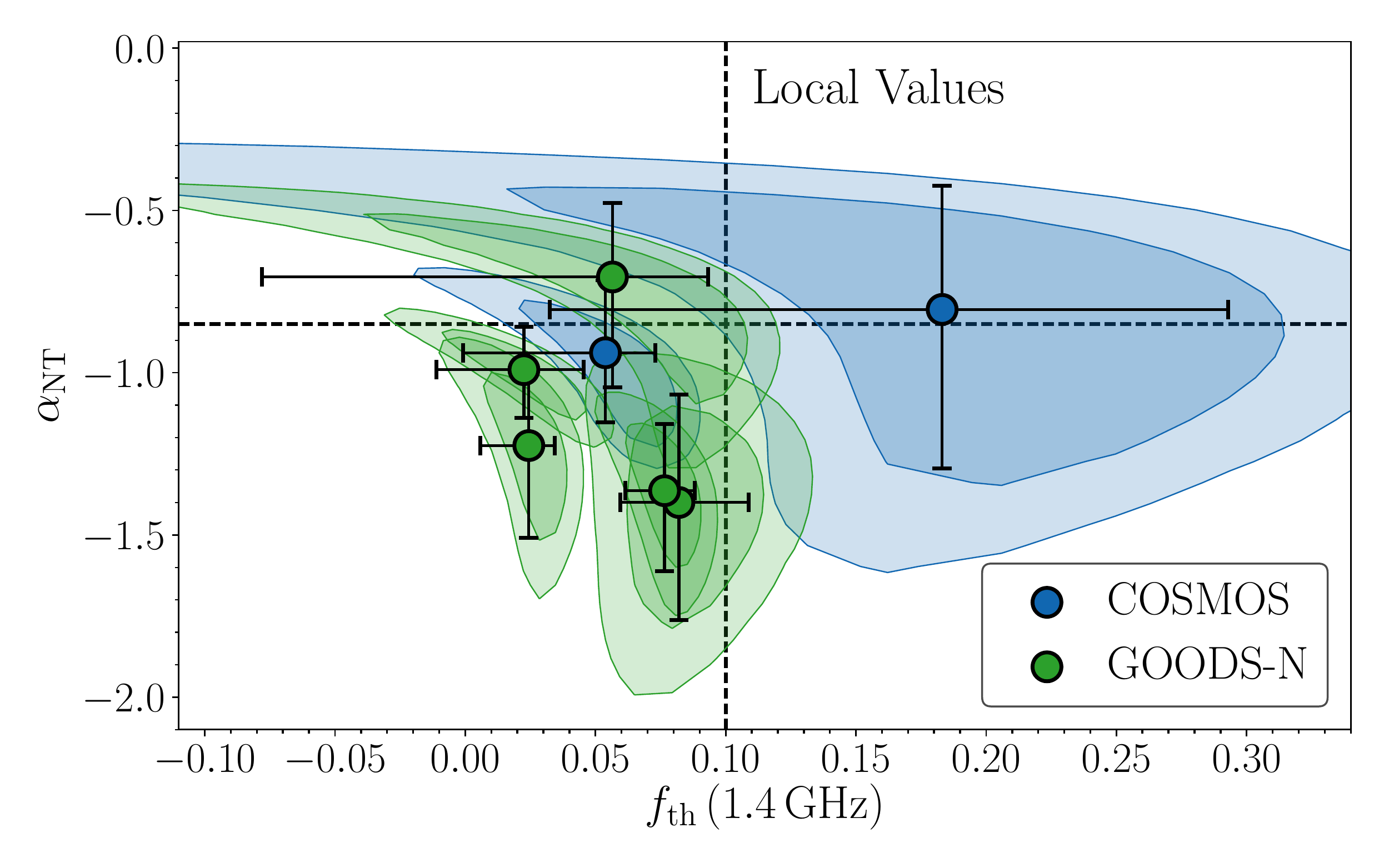}
    \caption{The fitted synchrotron spectral index versus the rest-frame 1.4\,GHz thermal fraction for the seven star-forming COLD$z$ continuum detections. The blue and green shaded regions indicate the covariance between the parameters, with the darker (lighter) shading indicating the $1\sigma$ ($2\sigma$) confidence regions. The datapoints show the corresponding one-dimensional uncertainties, and the typical local values are indicated through the dashed lines. The COLD$z$ galaxies are predominantly located in the lower left quadrant, indicating slightly steeper synchrotron spectral indices, and lower thermal fractions than what is typically observed locally.}
    \label{fig:spectral_params}
\end{figure*}

We show the radio spectra of the seven remaining star-forming sources across the COSMOS and GOODS-N fields in Figure \ref{fig:SFG_spectra}. All sources can be well-described by the combination of a synchrotron and free-free component, although for three sources (COS-3, GN-7, GN-11) -- while some contribution from free-free emission is preferred -- the fitted thermal fractions are consistent with zero within $1\sigma$. In turn, for these sources a single power-law representing synchrotron emission is sufficient to match the observed flux densities. We additionally emphasize that there is considerable covariance between the thermal fraction and synchrotron slope, and as such any quoted one-dimensional uncertainties are not fully representative of the multi-dimensional posterior distribution (Figure \ref{fig:spectral_params}). We therefore utilize these full posterior distributions in order to propagate the uncertainties into physical quantities such as free-free star-formation rates (Section \ref{sec:freefreeSFR}).

The fitted spectral parameters are presented in Figure \ref{fig:spectral_params} and Table \ref{tab:spectral_params}. Our bootstrapped median thermal fraction, scaled to 1.4\,GHz rest-frame as is common in the literature, equals $f_\text{th} = 0.06 \pm 0.03$, with a standard deviation of $\sigma=0.05$. None of the seven star-forming galaxies exhibit thermal fractions of $f_\text{th} \gtrsim 0.20$, indicating a fairly narrow distribution of $f_\text{th}$ at rest-frame 1.4\,GHz, even among a sample showing substantial variation in star-formation rates. Our average thermal fraction is slightly lower than the average value observed by \citet{tabatabaei2017} of $f_\text{th}=0.10$ for star-forming galaxies in the local Universe, though they report a large scatter of $\sigma = 0.09$. Additionally, the typical thermal fraction is similar to what was observed by \citet{niklas1997}, who determined $f_\text{th} = 0.08 \pm 0.01$ at 1\,GHz, with a scatter of $\sigma= 0.04$ across 74 local galaxies.

\begin{figure*}[!t]
    \centering
    \includegraphics[width=1.00\textwidth]{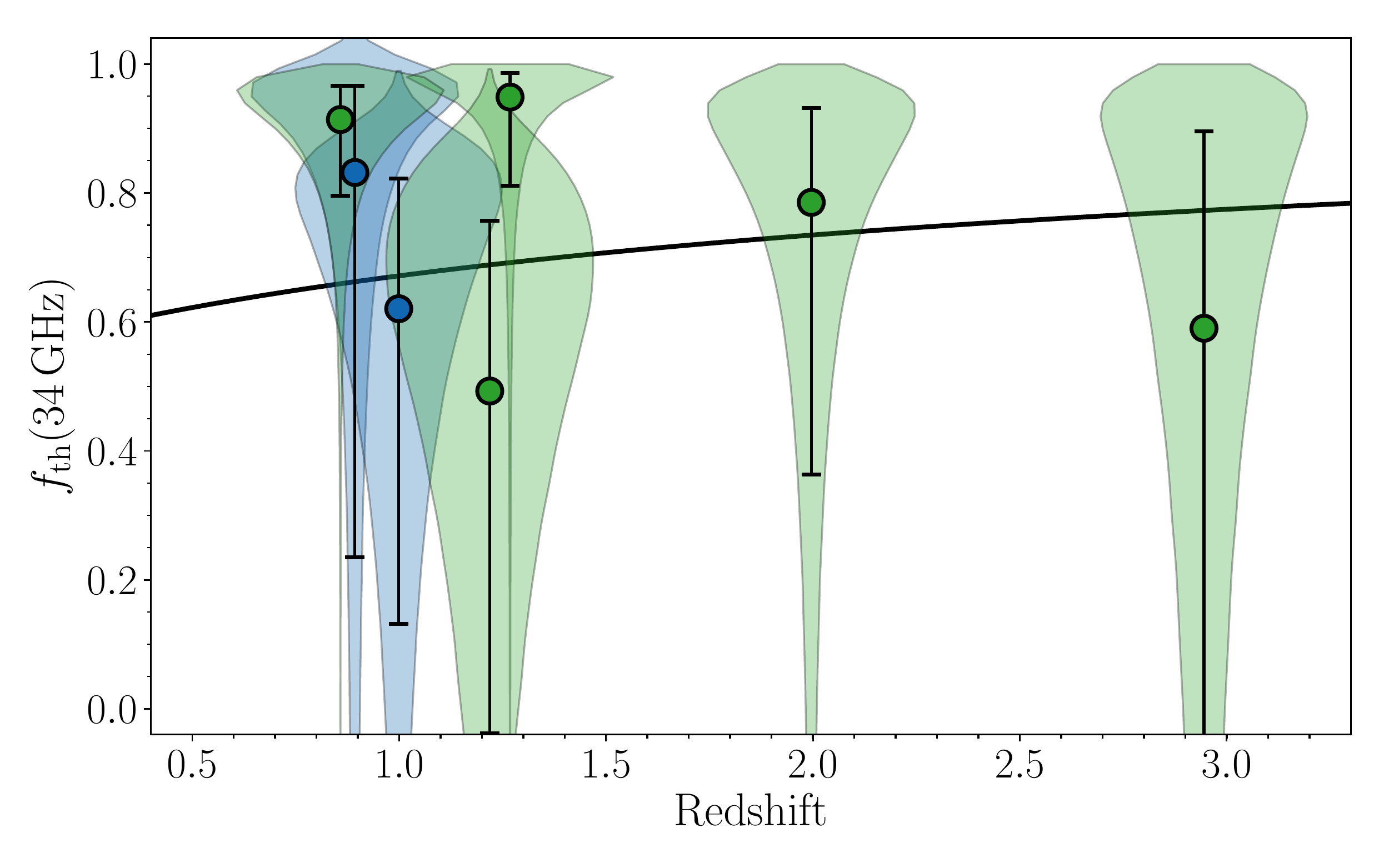}
    \caption{Violin diagrams of the thermal fraction at 34\,GHz (observed-frame, hence probing rest-frame $34\times(1+z)$\,GHz) as a function of redshift. The width of the shaded regions represents the probability distribution of the thermal fraction, while the datapoints indicate the median and 16-84$^\text{th}$ percentiles. The probability distributions of the thermal fractions are characterized by a typically high probability of having a large thermal fraction $f_\text{th} \gtrsim 0.7-0.8$, with a long tail extending towards lower values. The model from \citet{condon1992}, adopting $\alpha_\text{NT} = -0.85$ and $f_\text{th} = 0.10$ at 1.4\,GHz, is shown as the solid black line, and predicts similar high-frequency thermal fractions as observed across the star-forming galaxies.} \label{fig:fth_vs_freq}
\end{figure*}

We further determine an average thermal fraction at observed-frame 34\,GHz (i.e., probing $34\times(1+z)$\,GHz rest-frame) of $f_\text{th} = 0.78 \pm 0.07$, with a range of $f_\text{th} = 0.45 - 0.95$, and a scatter of $\sigma = 0.20$ (Figure \ref{fig:fth_vs_freq}). As such, we find that, even at rest-frame frequencies $\nu\gtrsim60\,$GHz, the radio spectrum is not fully dominated by thermal free-free emission, though we caution the uncertainties on the individual thermal fractions are large. We first compare these results with two local studies, both of which map free-free emission on sub-kpc scales. At a typical resolution of $\approx0.9\,$kpc, \citet{murphy2012} find an average thermal fraction across 103 star-forming regions of $f_\text{th} = 0.76$ at rest-frame 33\,GHz, with a scatter of $\sigma=0.24$. Extrapolating this value via the simple model from \citet{condon1992}, the typical thermal fraction at $\sim60\,$GHz is expected to be $\sim0.85-0.90$, which is slightly higher than the thermal fraction we find for high-redshift star-forming galaxies. In addition, \citet{linden2020} recently measured a typical thermal fraction at 33\,GHz of $f_\text{th} = 93\pm0.8\%$ across 118 star-forming complexes in local galaxies, at a resolution of $\approx0.2\,$kpc. Their typical thermal fraction is both higher than was determined by \citet{murphy2012}, and the values measured in this work. This, however, is not surprising, given that the thermal fractions presented in this work are integrated over the entire galaxy. As free-free emission is predominantly produced in star-forming regions, spatial variations in the thermal fraction across a galaxy are naturally expected, with the thermal fraction peaking in star-forming complexes. 

At high-redshift, no previous studies have directly targeted the free-free-dominated regime ($\nu\gtrsim30\,$GHz) in blindly selected galaxy samples, at a depth where star-forming galaxies are expected to dominate the radio population. However, particularly in bright dusty star-forming galaxies, some works have serendipitously detected high-frequency radio continuum emission, typically as a byproduct when targeting the CO(1-0) line. \citet{thomson2012} detect free-free emission in two $z\sim2.9$ lensed submillimeter galaxies, and determine thermal fractions of $f_\text{th}\sim0.3 - 0.4$ at 34\,GHz. Other studies of highly star-forming galaxies \citep{aravena2013,huynh2017} have additionally detected radio continuum emission at observed-frame $\sim30-35\,$GHz, but had to assume fixed synchrotron spectral indices due to a lack of ancillary data. Nevertheless, they estimate thermal fractions between $f_\text{th}\sim40-70\%$. Overall, these studies find thermal fractions that are broadly consistent with, albeit typically slightly lower than, what we determine for the star-forming sample detected in our non-targeted 34\,GHz observations.

Finally, we compare our results with the 10\,GHz pointing from \citet{murphy2017} in GOODS-N. They determine thermal fractions from $1.4-10\,$GHz spectral indices for $\sim25$ galaxies, under the assumption of a fixed synchrotron slope of $\alpha_\text{NT} = -0.85$. At a typical rest-frame frequency of $\nu\sim20\,$GHz, they find a median thermal fraction of $f_\text{th}\approx50\%$. Extrapolating this value to rest-frame 60\,GHz, this would imply a thermal fraction of $f_\text{th}\sim0.7$, similar to what we observe among the COLD$z$ star-forming sample.

We further determine a median non-thermal spectral index for the star-forming COLD$z$ sample of $\alpha_\text{NT} = -0.99_{-0.37}^{+0.19}$, with a standard deviation of $\sigma=0.25$. This typical value is consistent with the value observed by \citet{tabatabaei2017} for local star-forming galaxies of $\alpha_\text{NT} = -0.97 \pm 0.16$, but is slightly steeper than that of individual star-forming regions in NGC 6946, where \citet{murphy2011} find a typical value of $\alpha_\text{NT} = -0.81 \pm 0.02$. This is not surprising, as these observations directly target the acceleration sites of cosmic rays, where the spectrum should be flatter. However, our median synchrotron slope is additionally slightly steeper than the value obtained by \citet{niklas1997}, who determine an average $\alpha_\text{NT} = -0.83 \pm 0.02$ ($\sigma=0.13$) across 74 local galaxies. The slightly steeper non-thermal spectral index we find for the COLD$z$ sample may be the result of the higher rest-frame frequencies probed in this work, compared to the aforementioned local studies. Synchrotron cooling losses increase towards high rest-frame frequencies, and result in the steepening of the non-thermal spectral index (e.g., \citealt{thomson2019}). While our data do not have the constraining power to determine whether such spectral ageing occurs, as this requires additional sampling of the radio spectrum, such increased high-frequency losses would be fitted by a relatively steep synchrotron slope in our two-component model, and may plausibly contribute to our moderately lower value for $\alpha_\text{NT}$ compared to local studies.

Based on the spectral parameters we determine for the COLD$z$ sample, we calculate the rest-frame frequency $\nu_{50}'$ where the thermal fraction reaches 50\% via

\begin{align}
    \nu_{50}' = \nu' \left( \frac{1-f_\text{th}(\nu')}{f_\text{th}(\nu')} \right)^{-1/(\alpha_\text{NT} + 0.10)} \ .
\end{align}

\noindent While the uncertainties on this value are substantial for the individual star-forming galaxies due to its dependence on both the thermal fraction and synchrotron slope, we determine a median value of $\nu_{50}' = 11.5_{-7.4}^{+20.0}\,$GHz, with a standard deviation of $\sigma=5.9\,$GHz. This is slightly lower than, albeit still consistent with, the canonically assumed value of $\nu_{50}' \approx 25 - 30\,$GHz (e.g., \citealt{condon1992}), which is likely due to the relatively steep synchrotron slopes we are finding for the COLD$z$ star-forming galaxies.

While the differences between the spectral parameters determined for COLD$z$ and those observed in local star-forming galaxies are of minor statistical significance, a high-frequency selection of star-forming galaxies is likely to bias the sample towards having overall shallow radio spectra. Combining the $10-34\,$GHz spectral indices for the star-forming galaxies across COSMOS and GOODS-N, we find an average slope of $\alpha^{10}_{34} = -0.47 \pm 0.10$ ($\sigma=0.28$), which is substantially shallower than the canonical $\alpha=-0.70$ assumed at lower frequencies. Shallow spectra are naturally expected in the free-free dominated regime, in particular for young starburst galaxies. Such sources should exhibit large thermal fractions, as synchrotron emission lags the onset of a starburst by $\gtrsim 30\,$Myr (e.g., \citealt{bressan2002}), and the galaxies should hence be dominated by free-free emission across the entire radio spectrum. However, our sample is fully comprised of sources with typical or low thermal fractions ($f_\text{th} \lesssim 0.20$ at 1.4\,GHz), and rather steep synchrotron spectra ($\alpha_\text{NT} \lesssim -0.85$). This, in turn, indicates we are likely detecting relatively mature starbursts, as opposed to young star-forming galaxies. In particular, galaxies with a declining star-formation history are expected to exhibit only modest levels of free-free emission at low radio frequencies, and may show steepening of their synchrotron spectra towards higher frequencies. This scenario is qualitatively consistent with the spectral parameters we are finding for the COLD$z$ star-forming galaxies, and may therefore be typical for a high-frequency-selected sample. 

Alternatively, the relatively low thermal fractions observed for our star-forming sample at low frequencies may be the result of residual AGN contamination. An AGN will contribute additional synchrotron emission in excess of that arising from star-formation, while the overall contribution from free-free emission is mostly unaffected. In this case, one may expect typical synchrotron spectra, in combination with low thermal fractions. However, we find no strong evidence for systematic residual radio AGN activity based on the values of $q_\text{IR}$ of our star-forming galaxies, as we find that half of our sample falls onto or above the far-infrared/radio correlation for star-forming galaxies \citep{delhaize2017,algera2020a}.

\subsection{Free-Free Star-Formation Rates}
\label{sec:freefreeSFR}

\begin{figure*}[!t]
    \centering
    \hspace*{-0.1cm}\includegraphics[width=1.03\textwidth]{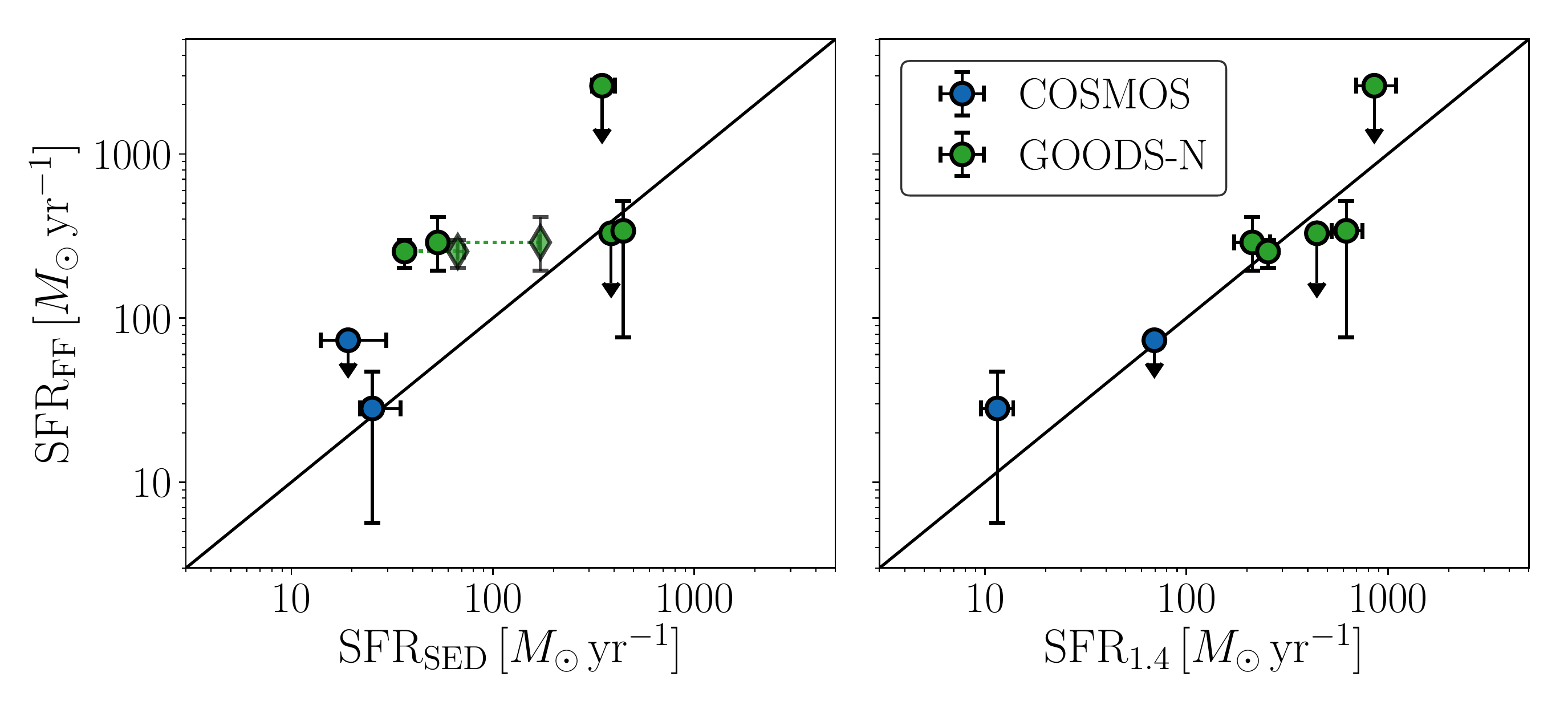}
    \caption{\textbf{Left:} A comparison of the star-formation rates obtained from free-free emission, versus those from {\sc{magphys}}. The one-to-one relation is shown through the solid black line, and the vertical errorbars represent the propagated uncertainties on the radio luminosity and thermal fraction. For three sources, we can only place upper limits on their free-free SFRs. The diamonds show the SFRs that we infer from SED-fitting when we instead convert the far-infrared luminosity into a star-formation rate, for the two sources where these values are discrepant (see text for details). \textbf{Right:} Free-free star-formation rates versus those obtained from radio synchrotron emission, adopting the FIRRC from \citet{delhaize2017}. Despite the uncertainties on $\text{SFR}_\text{FF}$ being large as a result of the low signal-to-noise at 34\,GHz, and the limited available sample size, we find that the free-free star-formation rates are in reasonable agreement with the values derived from SED-fitting and low-frequency radio emission.} \label{fig:freefree_sfrs}
\end{figure*}

Free-free emission is one of the most robust tracers of star-formation, as it constitutes a direct probe of the ionizing photons emitted by recently formed ($\lesssim10\,$Myr) massive stars. As such, it does not rely on reprocessed starlight, or emission produced by stellar remnants, such as, respectively, far-infrared and radio synchrotron emission. However, some caveats still apply, as carefully summarized in \citet{querejeta2019}. In particular, any ionizing photons absorbed by dust within the HII-region will not contribute to the ionization of hydrogen atoms, and as such will reduce the free-free luminosity at a fixed star-formation rate (e.g., \citealt{inoue2001,dopita2003}). Alternatively, if there is substantial leakage of ionizing photons, free-free emission will similarly be suppressed. Finally, since free-free emission is only sensitive to the massive end of the initial mass function, any variations in the IMF may substantially affect the calculated star-formation rates. We note that these caveats also apply to star-formation rates estimated using the Balmer lines (e.g., H$\alpha$, H$\beta$), with the clear advantage of free-free emission being that it is fully dust-insensitive on scales beyond the HII-region wherein the star-formation occurs.

With these caveats in mind, we now set out to calculate free-free star-formation rates for the seven star-forming galaxies detected at 34\,GHz. The calibration from \citet{murphy2012}, adapted to a Chabrier IMF, is given by

\begin{align}
\begin{split}
    \left( \frac{\text{SFR}_\text{FF}}{M_\odot\,\text{yr}^{-1}} \right) = 4.3 &\times 10^{-28} \left( \frac{T_e}{10^4\,\text{K}} \right)^{-0.45} \left( \frac{\nu}{\text{GHz}}\right)^{0.10} \\ 
    &\times \left( \frac{f_\text{th}(\nu) L_{\nu}}{\text{erg\,s}^{-1}\,\text{Hz}^{-1}}\right) \ .
\end{split}
\end{align}

\noindent Here $T_e$ is the electron temperature, which we assume to equal $T_e = 10^{4}\,$K. However, we note that our results are somewhat insensitive to the precise value adopted, given the modest exponent of $T_e^{-0.45}$. The star-formation rate is further directly proportional to the product of the thermal fraction and the radio luminosity, which we evaluate at a rest-frame frequency of $\nu = 1.4\,$GHz.

We derive typical free-free star-formation rates between $\text{SFR}\approx30-350\,M_\odot\,\text{yr}^{-1}$ for the 4/7 sources for which we have robustly constrained thermal fractions. The remaining sources instead have a thermal fraction that is consistent with zero within $1\sigma$, such that we can only provide upper limits on their free-free star-formation rates. We compare the free-free star-formation rates with those derived from {\sc{magphys}}, as well as from low-frequency radio synchrotron emission in Figure \ref{fig:freefree_sfrs}. For the latter, we adopt the far-infrared/radio correlation from \citet{delhaize2017}, which is suitable for the radio-detected star-forming population. While we have derived individual $q_\text{IR}$-values for the star-forming COLD$z$ sample in Section \ref{sec:multiwavelength}, we adopt a fixed far-infrared/radio correlation from the literature to ensure the synchrotron-derived star-formation rates are independent of those from far-infrared emission.  

Overall, we observe a reasonable agreement between the star-formation rates from free-free emission, and those from the more commonly adopted tracers we use for comparison. This likely implies that the various aforementioned caveats do not greatly affect our calculated SFRs. However, the correlation between the two radio-based tracers appears tighter than the one between the SFRs from free-free emission and SED-fitting, which is surprising as the timescale for far-infrared emission should more closely resemble that of FFE than synchrotron emission. Two sources in particular, GN-4 ($z=1.27$) and GN-9 ($z=0.87$) appear to have well-constrained free-free SFRs which exceed the ones from SED-fitting by a factor of $5.6\pm 0.1$ and $7.0 \pm 0.3$, respectively. If dust extinction within the HII-region, or leakage of ionizing photons were a concern, the free-free SFRs should be suppressed, in contrast to what we observe for these two sources. Instead, for GN-4 this offset is likely related to the SFR determined from SED-fitting. While the infrared SED of this galaxy is well-constrained (Figure \ref{fig:seds}), {\sc{magphys}} predicts that a substantial fraction of the infrared emission from GN-4 originates from an older stellar population, with the ratio between its infrared-based star-formation rate -- assuming the conversion from \citet{kennicutt1998} adjusted for a Chabrier IMF -- and the fitted SFR equaling $3.2 \pm 0.1$. Given that the SFRs from free-free emission and the 1.4\,GHz luminosity are in good agreement for GN-4, it is likely that the contribution from old stars to the FIR-luminosity is overestimated in the SED-fitting. 

A similar discrepancy can be seen between the free-free and infrared star-formation rates derived for GN-9. While this source too has a modest contribution to its dust luminosity from older stars, upon accounting for this, its ratio between the free-free and SED-fitted star-formation rates remains a factor of $3.8_{-0.3}^{+0.1}$. Instead, the radio flux densities of this source are likely to be boosted by the emission from an AGN. We find a far-infrared/radio correlation parameter of $q_\text{IR} = 1.99 \pm 0.04$ for GN-9 (Section \ref{sec:radioagn}), which implies it lies $\sim0.5\,$dex below the median value for radio-selected star-forming galaxies at its redshift of $z=0.87$ \citep{delhaize2017}, though still above the threshold we adopt for identifying radio AGN. Nevertheless, GN-9 is $\sim3.5\times$ radio-bright with respect to the median far-infrared/radio correlation, which fully accounts for the difference between its radio and SED-fitted star-formation rates. 

Interestingly, we can only place upper limits on the free-free star-formation rates for two of the brightest star-forming galaxies in GOODS-N ($\text{SFR}_\text{SED} \sim 400\,M_\odot\,\text{yr}^{-1}$), GN-7 and GN-11. While naively these galaxies might be expected to exhibit strong free-free emission, recent studies of starburst galaxies have found that their radio spectra might steepen towards higher frequencies \citep{tisanic2019,thomson2019}. This may be related either to a deficit in free-free emission, or to the steepening of their synchrotron spectra. However, these studies were limited to rest-frame frequencies $\nu \lesssim 20\,$GHz, and hence did not probe the regime where free-free emission is expected to dominate. While our study does directly target this high-frequency regime, with the current sampling of the radio spectrum we are unable to distinguish between a deficit in free-free emission, or more complex behavior in the synchrotron emission. Nevertheless, our lack of a robust detection of free-free emission in these strongly star-forming sources provides support for the existence of more complex radio spectra, which may be constrained by probing their radio emission at intermediate frequencies, using, for example, the VLA K-band (22\,GHz).

Taking the aforementioned caveats into account, our derived free-free star-formation rates are in good agreement with those from SED-fitting and the low-frequency far-infrared/radio correlation. While the uncertainties on the free-free SFRs remain large as a result of the typical faintness of star-forming galaxies at high rest-frame frequencies, our analysis indicates that a deep, 34-GHz selected sample, supplemented by deep ancillary radio observations, can be used to accurately constrain star-formation at high-redshift. This, in turn, will be possible for significantly larger galaxy samples in the future, with the increased sensitivity of next-generation radio facilities.

\renewcommand{\arraystretch}{1.45}
\begin{deluxetable*}{lcccccc}

\tabletypesize{\footnotesize}
\tablecolumns{7}
\tablewidth{\textwidth}
	
\tablecaption{Spectral parameters of the star-forming COLD$z$ sample}

\tablehead{
	\colhead{\textbf{ID}} &
	\colhead{$\nu'$\tablenotemark{a}} &
	\colhead{$f_\text{th}(1.4\,\text{GHz})$\tablenotemark{b}} &
	\colhead{$f_\text{th}(\nu'$)\tablenotemark{c}} &
	\colhead{$\alpha_\text{NT}$} &
	\colhead{$\text{SFR}_\text{FF}$\tablenotemark{d}} &
	\colhead{$\text{SFR}_{1.4}$}
    }

\startdata

& [GHz] & & & & [$\log M_\odot\,\text{yr}^{-1}$] & [$\log M_\odot\,\text{yr}^{-1}$] \\ \tableline \vspace{-0.3cm} \\

COLDz-cont-COS-1 & $64$ & $0.18_{-0.15}^{+0.11}$ & $0.82_{-0.71}^{+0.14}$ & $-0.81_{-0.49}^{+0.38}$ & $28_{-23}^{+19}$ & $12_{-2}^{+2}$ \\
COLDz-cont-COS-3 & $68$ & $0.05_{-0.05}^{+0.02}$ & $0.60_{-0.61}^{+0.22}$ & $-0.94_{-0.21}^{+0.22}$ & $53_{-54}^{+20}$ & $69_{-5}^{+5}$ \\
COLDz-cont-GN-4 & $77$ & $0.08_{-0.02}^{+0.03}$ & $0.95_{-0.14}^{+0.04}$ & $-1.40_{-0.36}^{+0.33}$ & $289_{-94}^{+123}$ & $213_{-40}^{+47}$ \\
COLDz-cont-GN-7 & $134$ & $0.06_{-0.13}^{+0.04}$ & $0.50_{-1.20}^{+0.38}$ & $-0.71_{-0.34}^{+0.23}$ & $1440_{-3530}^{+1181}$ & $857_{-161}^{+239}$ \\
COLDz-cont-GN-9 & $63$ & $0.08_{-0.02}^{+0.01}$ & $0.91_{-0.12}^{+0.05}$ & $-1.36_{-0.25}^{+0.21}$ & $254_{-52}^{+45}$ & $254_{-18}^{+19}$ \\
COLDz-cont-GN-11 & $75$ & $0.02_{-0.03}^{+0.02}$ & $0.45_{-0.74}^{+0.30}$ & $-0.99_{-0.15}^{+0.13}$ & $162_{-243}^{+166}$ & $444_{-32}^{+34}$ \\
COLDz-cont-GN-12 & $102$ & $0.02_{-0.02}^{+0.01}$ & $0.78_{-0.53}^{+0.16}$ & $-1.22_{-0.28}^{+0.24}$ & $342_{-260}^{+174}$ & $622_{-96}^{+127}$

\enddata

\tablenotetext{a}{The rest-frame frequency probed by the COLD$z$ 34\,GHz observations.}
\tablenotetext{b}{The thermal fraction at rest-frame 1.4\,GHz.}
\tablenotetext{c}{The thermal fraction at rest-frame frequency $\nu'$.}
\tablenotetext{d}{The uncertainties on the free-free SFRs are propagated from those on the thermal fraction and radio luminosity. The sources where these SFRs are consistent with zero within $1\sigma$ are plotted as upper limits in Figure \ref{fig:freefree_sfrs}.}

\label{tab:spectral_params}
\end{deluxetable*}
\renewcommand{\arraystretch}{1.0}

\section{Free-free Emission with the SKA and ngVLA}
\label{sec:nextgen}

The next large radio telescope to come online is the Square-Kilometer Array Phase 1 (SKA1), with SKA1-Mid set to cover a frequency range of $0.35 - 15\,$GHz. As such, observations with the highest-frequency band of the SKA1-Mid (at a central frequency of $\nu_c = 12.5\,$GHz) will start probing the regime where free-free emission dominates in star-forming galaxies at $z\gtrsim1$. A galaxy with $\text{SFR} = 10\,M_\odot\,\text{yr}^{-1}$ at $z=1$ will have a flux density of approximately $S_{12.5} \approx 1.5\,\mu$Jy in this band, assuming the FIRRC from \citet{delhaize2017}, and the calibration between star-formation and free-free emission from \citet{murphy2012}. This, in turn, requires $\sim15-20\,$hr of telescope time for a $5\sigma$ detection, based on the SKA1 sensitivity estimates from \citet{braun2019}. In order to robustly probe the free-free emission in such a modestly star-forming galaxy, additional sampling of its low-frequency radio spectrum is crucial. In particular, for a similar $5\sigma$ detection at $1.4$ and $6.7\,$GHz, a further $\sim10-15\,$hr of total telescope time is required. Given the $\sim4\times$ larger field-of-view at $6.7\,$GHz compared to at $12.5\,$GHz, a possible observing strategy for the detection of free-free emission in faint star-forming galaxies is to combine two single SKA1-Mid pointings at 1.4 and 6.7\,GHz with a five pointing mosaic at $12.5\,$GHz, covering the entire 6.7\,GHz field-of-view. With a total telescope time of $\sim 100\,$hr, this allows for the mapping of free-free emission in all $z\gtrsim1$ star-forming galaxies at $S_{12.5}\gtrsim1.5\,\mu$Jy across an area of $\sim120\,\text{arcmin}^2$. Adopting the \citet{bonaldi2019} simulations of the radio sky, developed specifically for the SKA1, a typical $\sim1100 - 1200$ galaxies are expected at $S_{12.5} \gtrsim 1.5\,\mu$Jy within this field-of-view. In particular, approximately $\sim68\pm2\%$ ($\sim12\pm1\%$) of this sample is expected to lie at a redshift $z\geq 1$ ($z\geq3$), allowing for the robust sampling of the free-free dominated regime. For comparison, the 200\,hr VLA COSMOS-XS survey \citep{vandervlugt2020}, reaches a similar depth to these template SKA1-Mid observations at 3\, and 10\,GHz, but covers a smaller area of $\sim30\,\text{arcmin}^2$. The increased survey speed of SKA1-Mid, therefore, allows for a $\gtrsim 8\times$ quicker mapping of free-free emission up to $\sim15\,$GHz, compared to the VLA. However, as the frequency coverage of SKA1-Mid is not fully optimized to directly probe the high-frequency radio emission in star-forming galaxies, significant synergy with the VLA remains, as it allows for the extension of the spectral coverage from the SKA1 to higher frequencies. 

The ngVLA \citep{murphy2018b,mckinnon2019}, however, is set to truly transform our understanding of the high-frequency radio spectrum in distant galaxies, and will allow for the usage of free-free emission as a high-redshift SFR-tracer on an unprecedented scale. The current sensitivity estimates \citep{butler2019} indicate that the ngVLA will attain a typical RMS of $\sigma\approx0.3\,\mu\text{Jy\,beam}^{-1}$ in one hour of Band four ($\nu = 20.5 - 34.0\,$GHz) observations. This, in turn, translates to a $5\sigma$ detection of a galaxy forming stars at $100\,M_\odot\,\text{yr}^{-1}$ at $z=2.5$, probing rest-frame $80\,$GHz, similar to the frequency range probed in this work for $z\approx2$ star-forming galaxies. However, with the expected improvement in sensitivity the ngVLA provides over the current VLA, high-redshift sources may more easily be targeted at relatively low observing frequencies, enabling wider surveys while still allowing for the free-free dominated regime to be probed. For example, \citet{barger2018} propose a survey at 8\,GHz (ngVLA Band two) to $0.2\,\mu\text{Jy\,beam}^{-1}$ across a large area of 1\,deg$^2$, which requires just an hour per pointing. By adopting a typical wedding cake strategy, deeper observations across a smaller area can further be used to target fainter star-forming galaxies. As an example, a star-forming galaxy of $\text{SFR}=25\,M_\odot\,\text{yr}^{-1}$ at $z=3$ ($z=5$) can in principle be detected at 8\,GHz in only $\sim3\,$hr ($\sim15\,$hr), modulo, of course, the large uncertainties on the typical thermal fraction in faint, star-forming sources, and the nature of the far-infrared/radio correlation in this population.


\section{CONCLUSIONS}
\label{sec:conclusion}

We have presented a deep continuum survey of the high-frequency radio sky with the Very Large Array, which probes the microJansky galaxy population at 34\,GHz. This regime has historically remained largely unexplored due to the relatively low survey speed of radio telescopes at these frequencies, as well as the expected faintness of sources. However, the high radio frequencies hold one of the most reliable tracers of star-formation, radio free-free emission, and as such are set to become a key area of study with next-generation radio facilities.

We employ deep observations at 34\,GHz from the COLD$z$ project \citep{pavesi2018,riechers2019,riechers2020} which cover the well-studied COSMOS ($10\,\text{arcmin}^2$) and GOODS-North ($50\,\text{arcmin}^2$) fields to a typical depth of $\sim1.5\,\mu\text{Jy\,beam}^{-1}$ and $\sim5.3\,\mu\text{Jy\,beam}^{-1}$, respectively. We perform source detection on the images down to a liberal $3\sigma$ detection threshold, aided by deep ancillary radio data across both fields, resulting in the detection of high-frequency continuum emission in eighteen galaxies. We cross-match these detections with additional deep radio observations at 1.4, 3 and 10\,GHz in the COSMOS field, as well as data at 1.4, 5 and 10\,GHz in GOODS-N. In addition, we leverage the wealth of multi-wavelength data across both fields to fully sample the SEDs of the galaxies from the X-ray to radio regime. The COLD$z$ continuum sample spans a redshift range of $z=0.50 - 5.30$, and lies at a median (mean) redshift of $z=1.12_{-0.15}^{+0.52}$ ($z=1.55_{-0.35}^{+0.41}$). The sample contains six sources at $z\geq2$, and includes the well-studied submillimeter galaxy AzTEC.3 at $z=5.3$. Our main findings are the following:

\begin{itemize}
    \item We present the first constraints on the radio number counts at 34\,GHz in the regime where star-forming galaxies dominate the radio population (Figure \ref{fig:numbercounts}), and find that these are in good agreement with lower-frequency number counts in the literature, both from observations \citep{vandervlugt2020} and simulations \citep{bonaldi2019}.
    
    \item We use the far-infrared/radio correlation to divide the 34\,GHz continuum sample into star-forming galaxies and active galactic nuclei (Figure \ref{fig:FIRRC}). In total, half of the sample (9 sources) shows AGN activity at radio wavelengths (Figure \ref{fig:agn_spectra}), while the radio emission of the remainder is consistent with being powered predominantly through star-formation. All but one of the faintest galaxies in our sample ($S_{34} \lesssim 20\,\mu$Jy) show radio emission of a star-forming origin, which is qualitatively consistent with the small fraction of radio AGN found in deep observations at lower frequencies (e.g., \citealt{algera2020b}). Two sources, including AzTEC.3 at $z=5.3$, likely have their continuum emission at 34\,GHz dominated by thermal emission from dust, leaving 7/18 sources ($\sim40\%$) of the sample with high-frequency radio continuum emission dominated by the combination of synchrotron and free-free emission.
    
    \item We use the wealth of ancillary radio data across the COSMOS and GOODS-N fields to construct radio spectra of the star-forming galaxies, covering four frequencies in the range $1.4 - 34\,$GHz (Figure \ref{fig:SFG_spectra}). We fit the radio spectra with a combination of free-free and synchrotron emission, and determine thermal fractions and non-thermal spectral indices for our sample (Figures \ref{fig:spectral_params} and \ref{fig:fth_vs_freq}), which are consistent with the values observed in local galaxies. We further determine free-free star-formation rates for seven star-forming galaxies, and find good agreement with those obtained from SED-fitting and the far-infrared/radio correlation (Figure \ref{fig:freefree_sfrs}).
    
\end{itemize}

With the 34\,GHz continuum data from the COLD$z$ survey, we have directly targeted free-free emission in faint star-forming sources at high redshift. While currently limited to a modest sample, next-generation radio facilities are set to significantly increase the number of galaxies for which the full radio spectrum is constrained, and will transform our understanding of high-frequency radio continuum emission in star-forming galaxies. However, combined with the wealth of ancillary data in the COSMOS and GOODS-N fields, the COLD$z$ observations already allow for a census of free-free emission in the typical star-forming population via a multi-frequency radio stacking analysis, which will be presented in a forthcoming publication (Algera et al. 2021, in preparation).


\section*{Acknowledgements}
H.S.B.\ Algera would like to thank A.\ de Graaff and D.\ Blanco for useful and constructive discussions. The authors thank Chris Carilli and Ian Smail for providing comments on the manuscript, and Fabian Walter for his help in the early stages of this project. The National Radio Astronomy Observatory is a facility of the National Science Foundation operated under
cooperative agreement by Associated Universities, Inc. H.S.B.A. and J.A.H. acknowledge support of the VIDI research programme with project number 639.042.611, which is (partly) financed by the Netherlands Organization for Scientific Research (NWO). D.R. acknowledges support from the National Science Foundation under grant numbers AST-1614213 and AST-1910107. D.R. also acknowledges support from the Alexander von Humboldt Foundation through a Humboldt Research Fellowship for Experienced Researchers. M.A. and this work have been supported by grants ``CONICYT+PCI+REDES 19019'' and ``CONICYT + PCI + INSTITUTO MAX PLANCK DE ASTRONOMIA MPG190030''.


\appendix

\section{Image Properties}
\label{app:radio_properties}
We discuss some of the properties of the COSMOS and GOODS-N 34\,GHz images below, including the completeness, and the level of flux boosting. In addition, we detail how we assign flux densities to the 34\,GHz continuum detections.

\subsection{Completeness}
\label{app:completeness}

We determine the completeness of the COSMOS and GOODS-N mosaics by inserting mock sources into the image, and extracting them via our regular source detection procedure (Section \ref{sec:sourcedetection}). The fraction of resolved sources at a resolution of $\sim2''$ at 10\,GHz -- similar to the resolution of our 34\,GHz observations -- is just $\sim10\%$ \citep{vandervlugt2020}, and is expected to be smaller at even higher frequencies, where sources are more compact \citep{murphy2017,thomson2019}. As such, we only include unresolved mock sources in our completeness analysis. Sources are inserted in the maps uncorrected for the primary beam (PB), as in this case it is straightforward to incorporate the incompleteness due to the decreased PB-sensitivity by inserting sources with a true flux density $S_\nu$ as $A(r)\times S_\nu$, where $A(r)$ represents the primary beam sensitivity at position $r$ within the mosaic. 

\begin{figure*}
    \centering
    \includegraphics[width=0.495\textwidth]{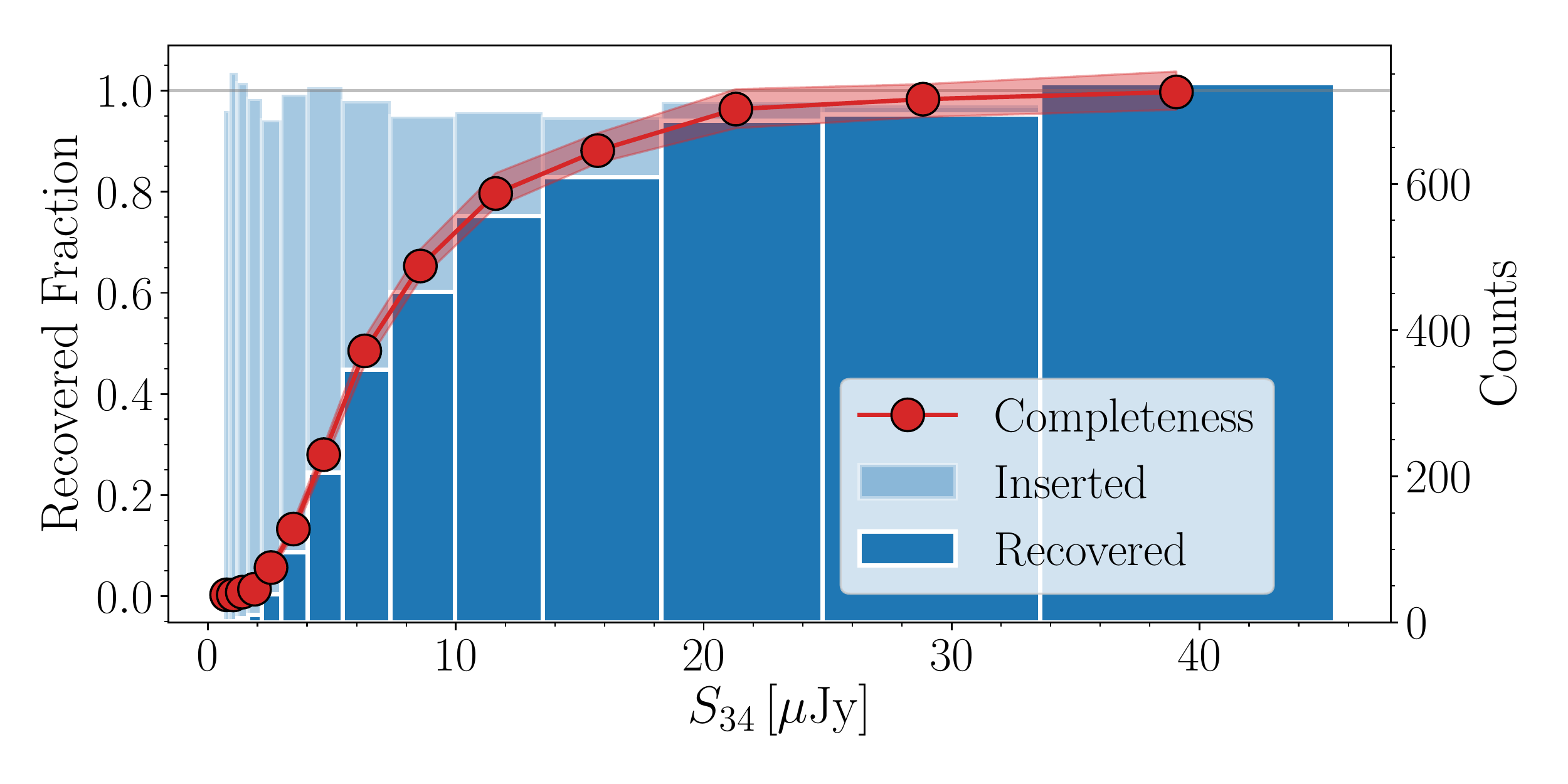}
    \includegraphics[width=0.495\textwidth]{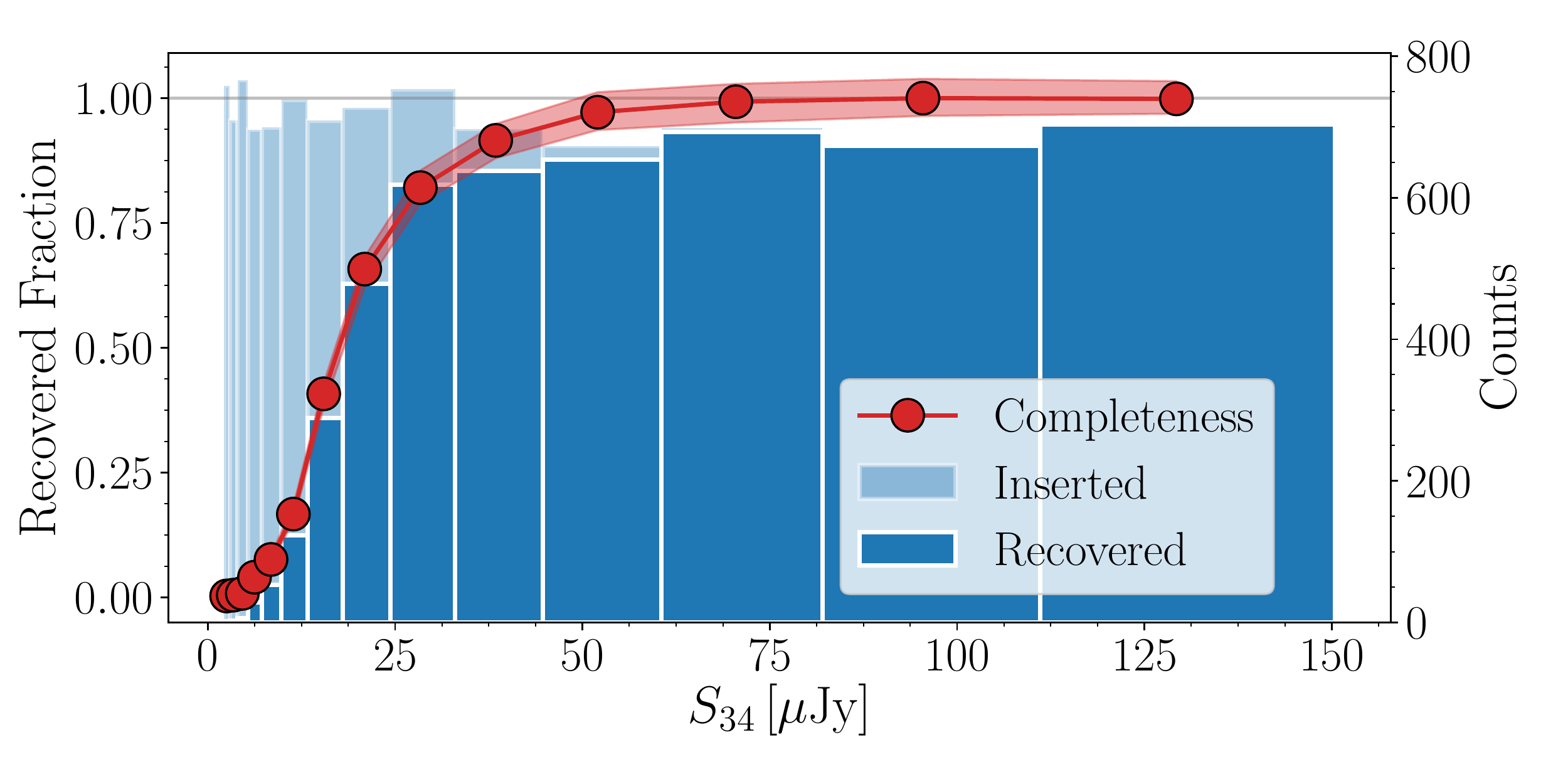}
    \caption{\textbf{Left:} Completeness analysis for the COSMOS field, showing the fraction of inserted mock sources we recover as a function of their flux density. The red datapoints and shaded region indicate the measured completeness and confidence region, respectively. \textbf{Right:} Completeness analysis for GOODS-North. Both panels take into account the variation in the primary beam sensitivity across the mosaic, and as such the total completeness correction constitutes the combination of missing sources due to a decreased primary beam sensitivity, and local noise properties within the mosaics.}
    \label{fig:completeness}
\end{figure*}

We randomize source positions within both mosaics, above $A(r) \geq 0.20$, and draw flux densities from a powerlaw distribution to ensure low flux densities -- where incompleteness will be the largest -- are amply sampled. For COSMOS, we insert 50 mock sources each run, for a total of 200 runs. As the GOODS-N mosaic is substantially larger, we instead insert 100 mock sources per run, for 100 runs total. In both cases, mock sources are required to be $2.5$ beam sizes away from both real sources and other mock sources. We then repeat the source detection procedure described in Section \ref{sec:sourcedetection}, and cross-match the recovered sources to the inserted ones, using a matching radius of $0\farcs7$. We record their inserted flux density, as well as their recovered peak and integrated flux densities. We define the completeness in a given flux density bin $i$ as $C_i = N_{\text{rec},i} / N_{\text{ins},i}$, that is, as the ratio of the number of inserted and recovered mock sources with a flux density that falls within the $i$-th bin. We determine the corresponding uncertainty via a bootstrap analysis, whereby we resample from the inserted flux densities, with replacement, and determine for each flux density bin the fraction of this sample that was recovered in our source detection procedure. The uncertainty then represents the $16^\text{th}-84^\text{th}$ percentile of the bootstrapped completeness analyses. We show the completeness in both the COSMOS and GOODS-N mosaics in Figure \ref{fig:completeness}. 

In the COSMOS field, we reach 50 and 80 per cent completeness at flux densities of $S_{34} = 7.3\,\mu\text{Jy\,beam}^{-1}$ ($\approx 6\sigma$, where $\sigma$ represents the typical RMS in the map) and $S_{34} = 13.3\,\mu\text{Jy\,beam}^{-1}$ ($\approx10\sigma$), respectively. In GOODS-North, we reach completeness fractions of 50 and 80 per cent at $S_{34} = 19.6\,\mu\text{Jy\,beam}^{-1}$ ($\approx 4\sigma$) and $S_{34} = 29.3\,\mu\text{Jy\,beam}^{-1}$ ($\approx 6\sigma$), respectively. These differences may be explained by the non-uniform exposure map of the GOODS-N mosaic, allowing for faint sources to still be detected in a small portion of the mosaic.

\subsection{Peak versus Integrated Fluxes}
\label{app:bondi}
\begin{figure*}
    \centering
    \includegraphics[width=0.495\textwidth]{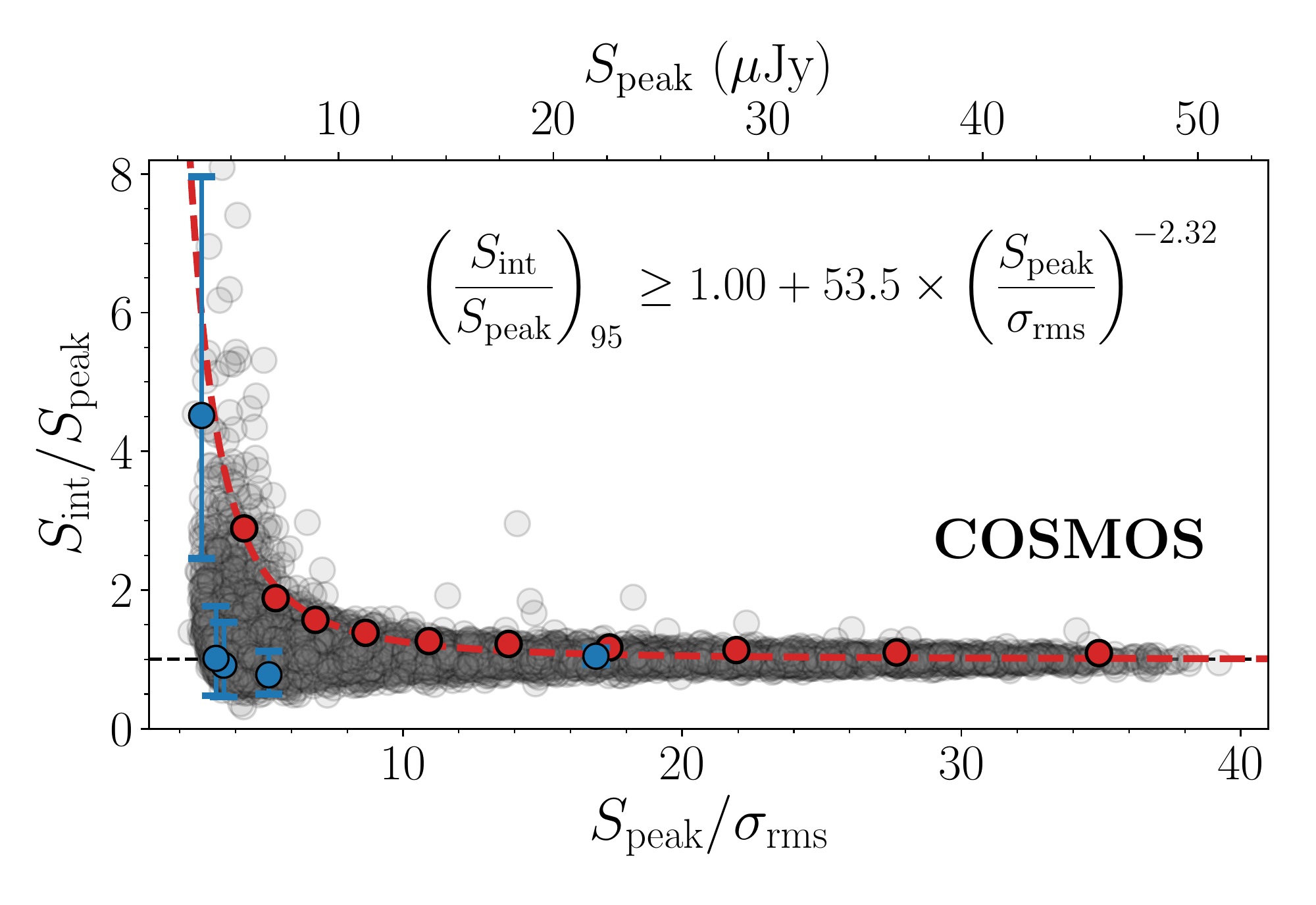}
    \includegraphics[width=0.495\textwidth]{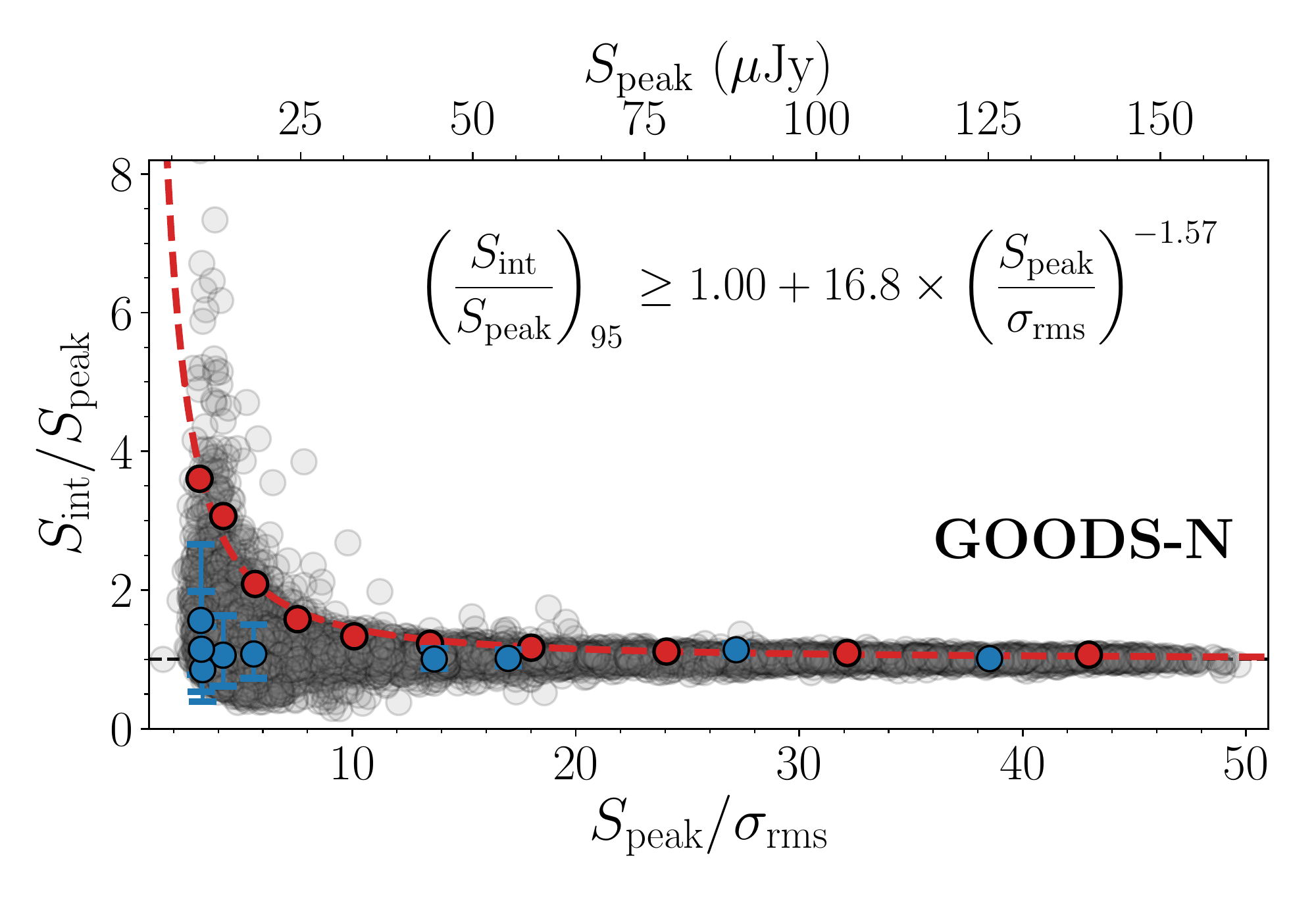}
    \caption{The ratio of integrated to peak flux density as a function of the peak SNR for unresolved mock sources inserted into the COSMOS and GOODS-N fields. Both panels show a power-law fit to the upper 95th percentile of each bin in SNR (red points) via the red, dashed line. Sources below this line are taken to be unresolved, and for these the peak brightness is adopted. The robust sources detected in both fields are shown in blue. A single source in GOODS-N is consistent with being resolved, whereas the remaining continuum detections are unresolved. Since the mock sources are inserted into the mosaic uncorrected for the primary beam, the RMS is highly uniform, and hence the SNR can be mapped into a peak brightness (upper horizontal axis).}
    \label{fig:bondi}
\end{figure*}

In order to assign a flux density to the detected radio sources, we need to establish if they are resolved. Our observations have a typical beam-size of $\sim2\farcs5$, and as such, we expect most sources to be unresolved, based on the typical (sub-)arcsecond radio sizes of star-forming galaxies at low-frequencies \citep{cotton2018,jimenez-andrade2019}, and the finding that these sources are more compact at higher frequencies \citep{murphy2017,thomson2019}. To verify this, we use our runs of inserted mock sources, which by construction are unresolved, and compare their peak and integrated flux densities as a function of signal-to-noise. We show the results in Figure \ref{fig:bondi}, for both the COSMOS and GOODS-N fields. We divide the results into logarithmically spaced bins in SNR, and determine the integrated/peak ratio encompassing 95 percent of sources per bin (following, e.g., \citealt{bondi2008,vandervlugt2020}). These percentiles are fitted with a power-law, with the region below the best fit defining the limiting integrated/peak flux density where sources are taken to be unresolved. It is clear that at modest signal-to-noise, $\text{SNR} \lesssim 5$, sources can have $S_\text{int} / S_\text{peak} \gtrsim 2$, despite being unresolved. This is a result of nearby noise peaks elongating the source, or throwing off the fitting, which mostly affects the integrated flux density. We find that all sources, with the exception of one bright ($\text{SNR} \sim 25$) detection in GOODS-N, show an integrated/peak ratio which is consistent with the source being unresolved. As such, we adopt the integrated flux density for this single source, and the peak brightness for the rest.

\subsection{Flux Boosting}
\label{app:fluxboosting}

\begin{figure*}
    \centering
    \includegraphics[width=0.495\textwidth]{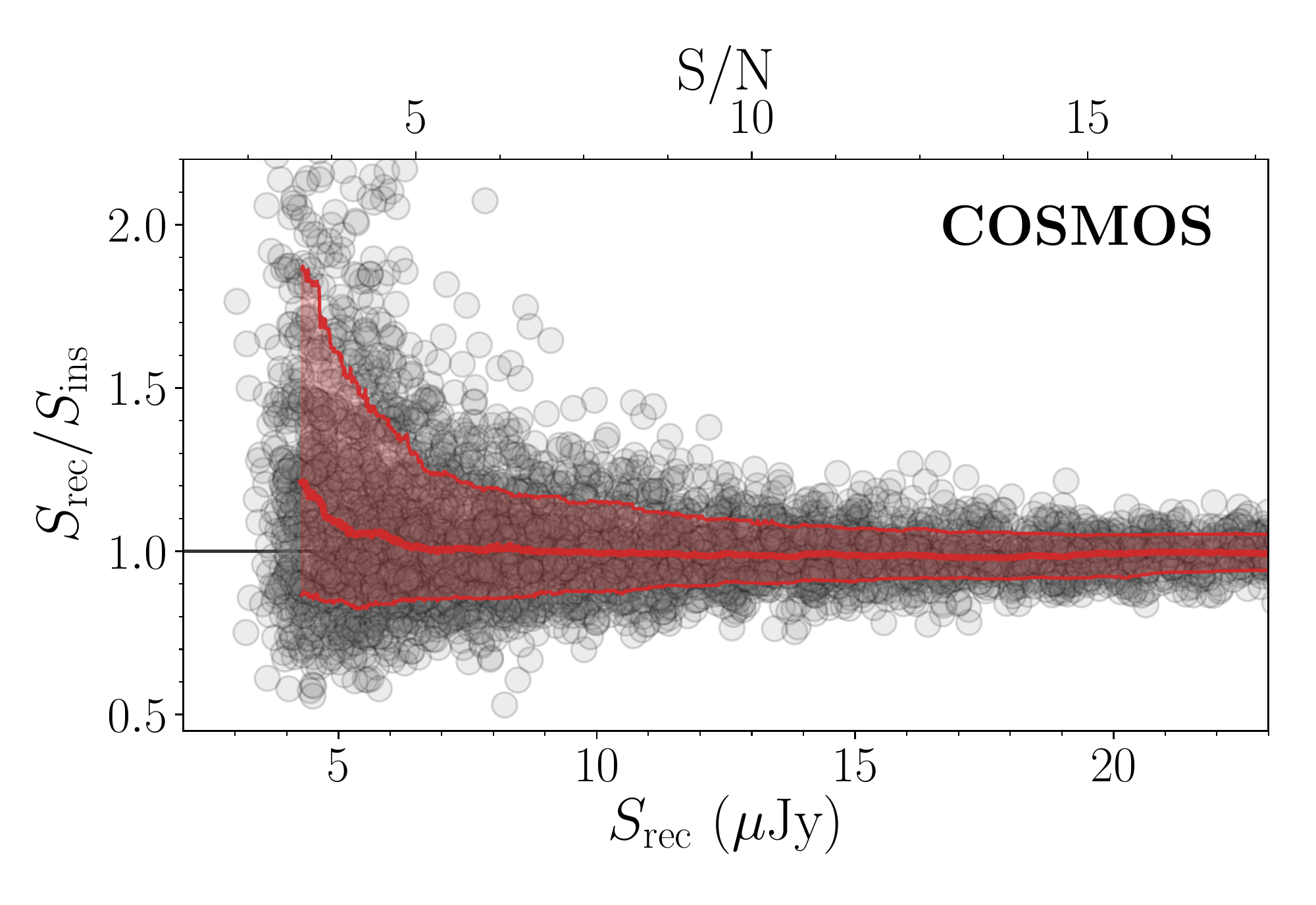}
    \includegraphics[width=0.495\textwidth]{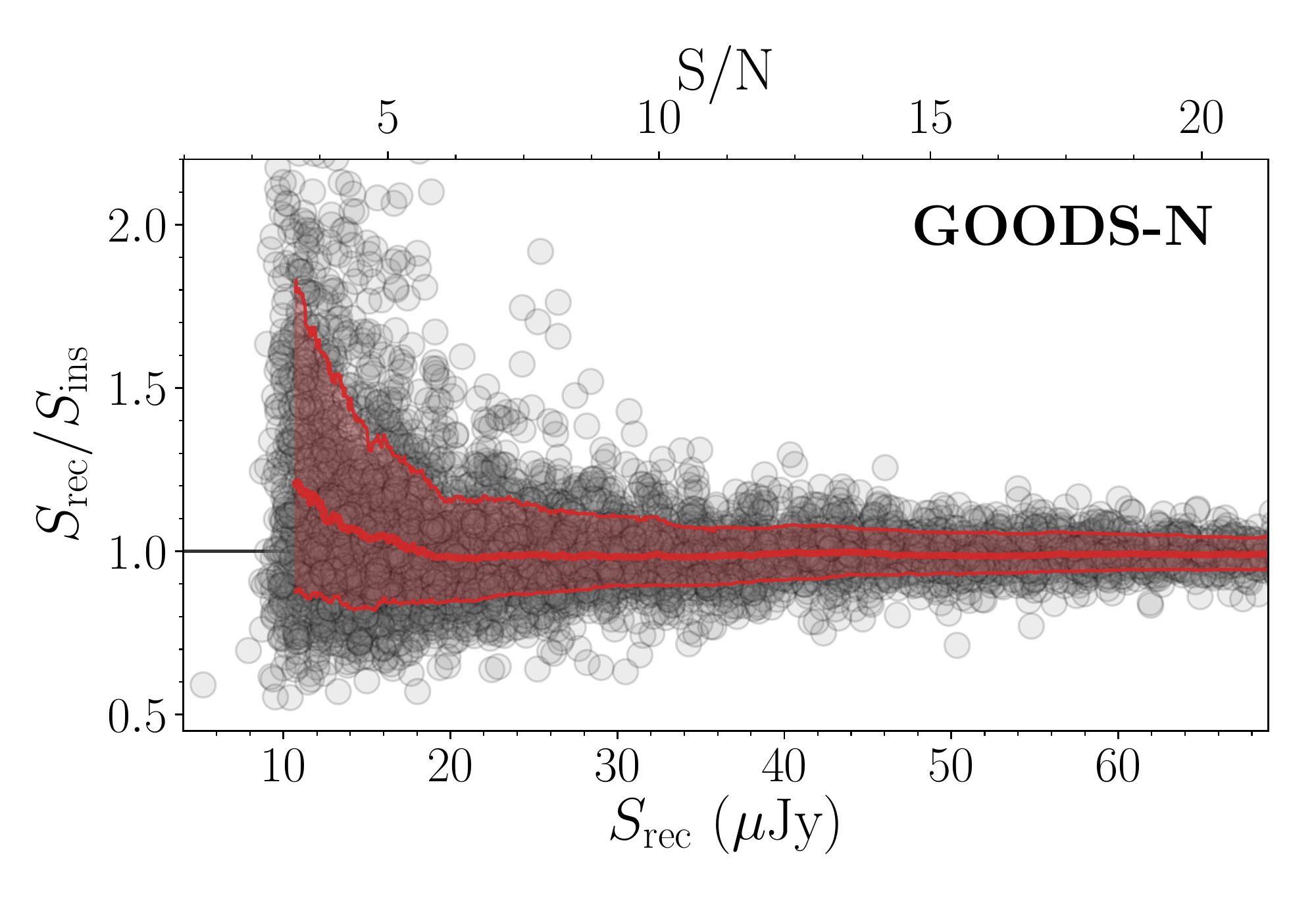}
    \caption{The ratio of recovered and inserted peak flux density as a function of recovered flux density, for unresolved mock sources inserted into the COSMOS and GOODS-N 34\,GHz maps. A running median and the corresponding $16-84^\text{th}$ percentile spread are indicated through the red line and shaded area. While for both fields the median level of flux boosting is negligible at $\text{SNR} \gtrsim 5$, the correction reaches $\sim20\%$ at $\text{SNR}\approx3$.}
    \label{fig:fluxboosting}
\end{figure*}

At low signal-to-noise, the peak brightness may be `boosted' as a result of noise properties in the image. To establish whether this is affecting the flux densities of our 34 GHz detections, we compare the recovered and inserted flux densities of our mock source analysis. The results are shown in Figure \ref{fig:fluxboosting}, for both COSMOS and GOODS-N. At $\text{SNR} \gtrsim 5$, the median ratio of recovered to inserted flux density is consistent with unity, with a spread of less than $\lesssim20\%$. At $\text{SNR} \lesssim 5$, which is the typical signal-to-noise at which the faintest 34\,GHz sources are detected, the level of flux boosting steadily increases, with a typical correction of $\sim20\%$ at $\text{SNR} \approx 3$ in either field. In this low-SNR regime, the typical spread on the ratio of recovered and inserted flux densities similarly increases strongly. Following, e.g., \citet{stach2019}, we correct the flux densities of the sources detected at 34\,GHz by the median level of flux boosting at their observed SNR. The uncertainty on the corrected flux density includes the propagated bootstrapped error on the median.

\section{Spectral Energy Distributions}
\label{app:seds}
\begin{figure*}[!t]
    \centering
    \vspace*{-1cm}
    \hspace*{-0.5cm}\includegraphics[width=0.98\textwidth]{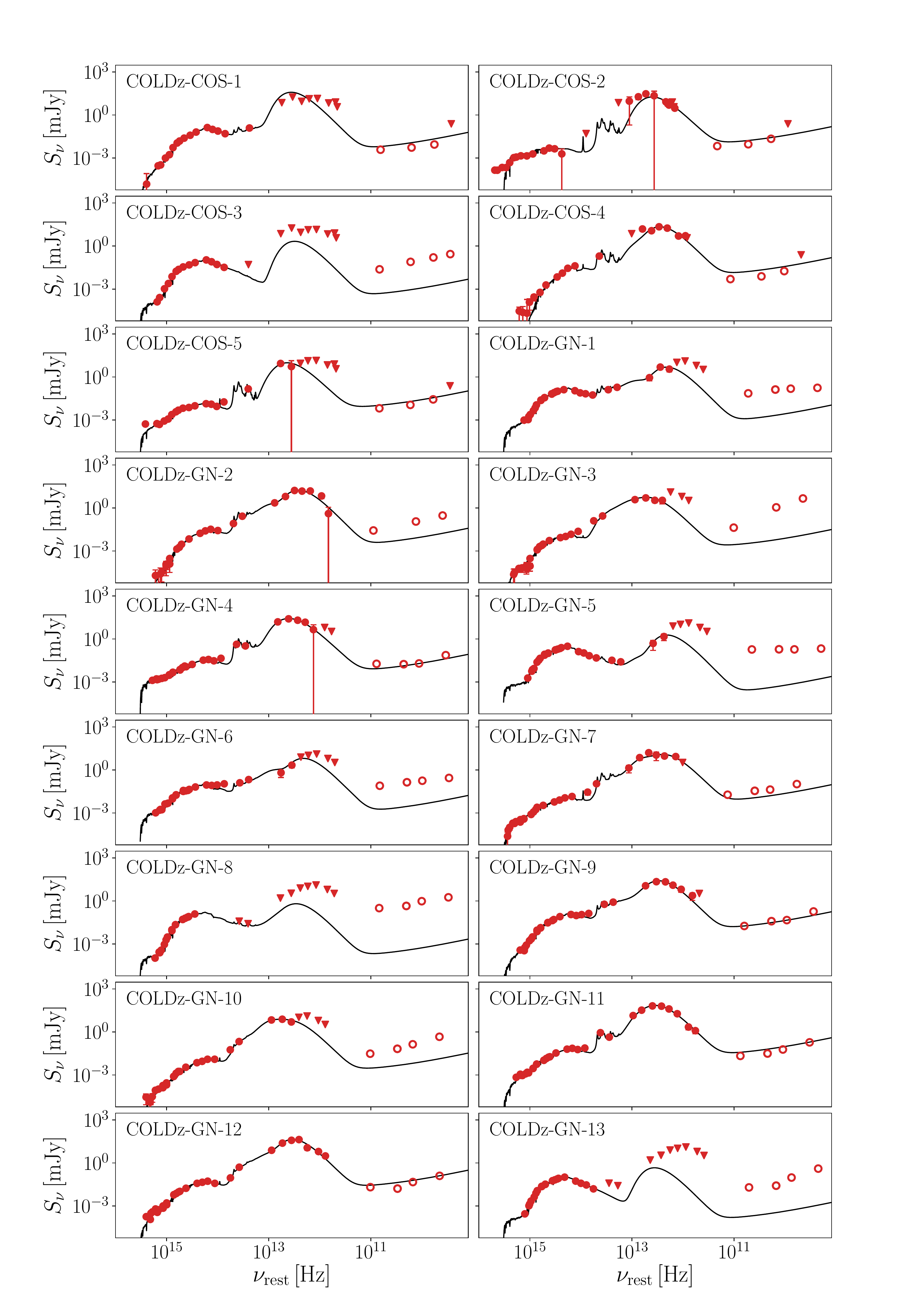}
    \caption{The spectral energy distributions of the 18 COLD$z$ continuum detections. All datapoints are shown in red, with triangles indicating upper limits and open symbols indicating the radio observations, which are not fitted. In half of the sources, the measured radio flux densities lie well above the extrapolated value from {\sc{magphys}}, indicative of additional emission from a radio AGN.}
    \label{fig:seds}
\end{figure*}

We show the full spectral energy distributions for the eighteen COLD$z$ continuum detections, fitted with {\sc{magphys}}, in Figure \ref{fig:seds}, and present their physical parameters in Table \ref{tab:magphys}.

\renewcommand{\arraystretch}{1.25}
\begin{deluxetable*}{lcccccc}

\tabletypesize{\footnotesize}
\tablecolumns{7}
\tablewidth{\textwidth}
	
\tablecaption{Physical Properties of the 34\,GHz-selected COLD$z$ sample}

\tablehead{
	\colhead{\textbf{ID}} &
	\colhead{$L_\text{IR}$} &
	\colhead{$\text{SFR}$\tablenotemark{1}} &
	\colhead{$M_*$} &
	\colhead{$L_{1.4}$\tablenotemark{2}} &
	\colhead{$q_\text{IR}$} &
	\colhead{\textbf{AGN}\tablenotemark{3}}
    }

\startdata

& [$\log L_\odot$] & [$\log M_\odot\,\text{yr}^{-1}$] & [$\log M_\odot$] & [$\log \text{W\,Hz}^{-1}$] & \\ \tableline \vspace{-1.0ex} \\

COLDz-cont-COS-1 & $11.75_{-0.09}^{+0.09}$ & $1.40_{-0.06}^{+0.14}$ & $11.39_{-0.06}^{+0.09}$ & $22.51_{-0.08}^{+0.08}$ & $3.25_{-0.12}^{+0.12}$ & 0 \\
COLDz-cont-COS-2 & $13.00_{-0.03}^{+0.01}$ & $2.80_{-0.03}^{+0.01}$ & $10.78_{-0.03}^{+0.02}$ & $24.83_{-0.06}^{+0.06}$ & $2.17_{-0.06}^{+0.06}$ & 0 \\
COLDz-cont-COS-3 & $10.70_{-0.51}^{+0.19}$ & $0.01_{-0.34}^{+0.17}$ & $11.16_{-0.01}^{+0.09}$ & $24.04_{-0.04}^{+0.04}$ & $0.67_{-0.49}^{+0.20}$ & 1 \\
COLDz-cont-COS-4 & $12.54_{-0.03}^{+0.10}$ & $2.21_{-0.40}^{+0.21}$ & $11.93_{-0.19}^{+0.03}$ & $24.01_{-0.06}^{+0.06}$ & $2.56_{-0.08}^{+0.10}$ & 0 \\
COLDz-cont-COS-5 & $11.38_{-0.16}^{+0.17}$ & $1.28_{-0.14}^{+0.19}$ & $10.23_{-0.08}^{+0.05}$ & $23.32_{-0.03}^{+0.03}$ & $2.08_{-0.17}^{+0.17}$ & 0 \\
COLDz-cont-GN-1 & $10.66_{-0.03}^{+0.07}$ & $0.31_{-0.11}^{+0.08}$ & $11.04_{-0.03}^{+0.05}$ & $23.17_{-0.02}^{+0.02}$ & $1.51_{-0.04}^{+0.06}$ & 1 \\
COLDz-cont-GN-2 & $12.12_{-0.02}^{+0.02}$ & $1.84_{-0.12}^{+0.07}$ & $11.47_{-0.09}^{+0.11}$ & $24.63_{-0.05}^{+0.13}$ & $1.49_{-0.13}^{+0.05}$ & 1 \\
COLDz-cont-GN-3 & $12.27_{-0.01}^{+0.01}$ & $1.92_{-0.01}^{+0.01}$ & $11.01_{-0.01}^{+0.01}$ & $26.18_{-0.02}^{+0.02}$ & $0.10_{-0.02}^{+0.02}$ & 1 \\
COLDz-cont-GN-4 & $12.23_{-0.01}^{+0.01}$ & $1.73_{-0.01}^{+0.01}$ & $10.64_{-0.01}^{+0.01}$ & $23.86_{-0.09}^{+0.09}$ & $2.37_{-0.09}^{+0.09}$ & 0 \\
COLDz-cont-GN-5 & $9.34_{-0.01}^{+0.21}$ & $-0.88_{-0.10}^{+0.17}$ & $10.83_{-0.06}^{+0.09}$ & $22.75_{-0.02}^{+0.02}$ & $0.62_{-0.04}^{+0.18}$ & 1 \\
COLDz-cont-GN-6 & $11.23_{-0.10}^{+0.04}$ & $1.11_{-0.11}^{+0.01}$ & $11.06_{-0.02}^{+0.05}$ & $23.93_{-0.02}^{+0.02}$ & $1.30_{-0.10}^{+0.05}$ & 1 \\
COLDz-cont-GN-7 & $12.60_{-0.07}^{+0.04}$ & $2.54_{-0.05}^{+0.07}$ & $10.81_{-0.07}^{+0.03}$ & $24.72_{-0.09}^{+0.10}$ & $1.88_{-0.12}^{+0.11}$ & 0 \\
COLDz-cont-GN-8 & $10.38_{-0.04}^{+1.31}$ & $-0.15_{-0.13}^{+1.01}$ & $11.56_{-0.12}^{+0.01}$ & $24.83_{-0.03}^{+0.03}$ & $-0.40_{-0.09}^{+1.25}$ & 1 \\
COLDz-cont-GN-9 & $11.82_{-0.01}^{+0.03}$ & $1.56_{-0.01}^{+0.02}$ & $11.23_{-0.01}^{+0.14}$ & $23.85_{-0.03}^{+0.03}$ & $1.99_{-0.03}^{+0.04}$ & 0 \\
COLDz-cont-GN-10 & $12.37_{-0.01}^{+0.01}$ & $2.34_{-0.01}^{+0.01}$ & $10.59_{-0.01}^{+0.06}$ & $25.11_{-0.02}^{+0.02}$ & $1.26_{-0.02}^{+0.02}$ & 1 \\
COLDz-cont-GN-11 & $12.58_{-0.01}^{+0.01}$ & $2.59_{-0.01}^{+0.01}$ & $11.01_{-0.01}^{+0.01}$ & $24.17_{-0.03}^{+0.03}$ & $2.42_{-0.03}^{+0.03}$ & 0 \\
COLDz-cont-GN-12 & $12.78_{-0.02}^{+0.01}$ & $2.65_{-0.01}^{+0.02}$ & $11.59_{-0.12}^{+0.01}$ & $24.46_{-0.07}^{+0.08}$ & $2.32_{-0.08}^{+0.07}$ & 0 \\
COLDz-cont-GN-13 & $9.43_{-0.15}^{+0.10}$ & $-1.00_{-0.08}^{+0.11}$ & $10.77_{-0.03}^{+0.02}$ & $23.62_{-0.03}^{+0.03}$ & $-0.19_{-0.15}^{+0.10}$ & 1

\enddata

\tablenotetext{1}{The star-formation rate as determined via {\sc{magphys}}.}
\tablenotetext{2}{The radio luminosity at rest-frame 1.4\,GHz is computed using the closest observed-frame flux density (1.4 or 3\,GHz) and the measured radio spectral index.}
\tablenotetext{3}{Sources are identified as radio AGN based on their offset from the far-infrared/radio correlation (Section \ref{sec:radioagn}).}

\label{tab:magphys}
\end{deluxetable*}
\renewcommand{\arraystretch}{1.0}

\bibliographystyle{apj}
\bibliography{main}

\end{document}